
\input phyzzx

\input epsf
\pubnum{OU-HET-211}
\date{June 1995}
\titlepage
\title{ Motions of the String Solutions in the XXZ Spin Chain under
 a Varying Twist  }
\author{\rm  N. Fumita, H. Itoyama and T. Oota
\foot{  This work is supported in part by a Grant-in-Aid for Scientific
 Research  ($07640403$) from the Ministry of Education, Japan.
e-mail:
 itoyama@funpth.phys.sci.osaka-u.ac.jp    toota@funpth.phys.sci.osaka-u.ac.jp }
}
\address{\sl Department of Physics,   Graduate School of Science,
  Osaka University, Toyonaka, Osaka 560, Japan}

\vskip 3cm

\centerline{}
\centerline{}
\centerline{}
\centerline{}

\abstract{ We determine  the motions of the roots of the
 Bethe ansatz equation for the ground state
 in the XXZ spin chain under a varying twist angle.
   This is done  by   analytic as well as  numerical
 study at a finite size system. In the attractive critical regime $ 0<
 \Delta <1 $,
 we reveal intriguing motions of  strings  due to
 the finite size corrections to the length of the strings: in
 the case of two-strings, the roots collide  into the branch points
  perpendicularly to the imaginary axis, while in the case of three- strings,
 they fluctuate around the center of the string.
  These are successfully generalized to the case of  $n$-string.
  These results are used to determine the final configuration
 of the momenta as well as  that of
 the phase shift functions.
 We obtain these as well as the period and the Berry phase
 also in the regime
 $ \Delta \leq -1$, establishing the continuity of the previous results
  at $ -1 < \Delta < 0 $ to this regime.
 We  argue  that the Berry phase
   can be used as a measure of the statistics of the  quasiparticle
 ( or the bound state) involved in the process.}

\endpage

\REF\BETHE{H. A. Bethe, {\sl Z. Physik}{\bf 71}(1931) 205.}

\Ref\HU{L. Hulthen, {\sl Ark. Mat. Astron. Fys.} {\bf 26A}11 (1938) 1.}

\REF\YY{C. N. Yang and C. P. Yang, {\sl Phys. Rev} {\bf 147} (1966) 303; {\bf
150} (1966) 321, 327; {\bf 151} (1966) 258.}

\REF\L{E. H. Lieb, {\sl Phys. Rev.} {\bf 162} (1967) 162; {\sl Phys. Rev.
Lett.} {\bf 18} (1967) 1046; {\bf 19} (1967) 108.}

\REF\G{M. Gaudin, {\sl Phys. Rev. Lett} {\bf 26} (1971) 1301.}

\REF\T{M. Takahashi, {\sl Prog. Theor. Phys.} {\bf 50} (1973) 1519;
 {\bf 51} (1974) 1348.}

\REF\KOR{V. E. Korepin, {\sl Commun. Math. Phys.} {\bf 86}(1982) 391.}

\REF\ABB{F. Alcaraz, M. Barber and M. Batchelor, {\sl Phys. Rev. Lett.}
 {\bf 58} (1987)771.}

\REF\PT{S.V. Pokorski and A.M. Tsvelick, {\sl Sov. Phys. JETP} {\bf 66}
  (1987)1275.}

\REF\PS{V. Pasquier and H. Saleur, {\sl Nucl. Phys.} {\bf B330} (1990) 523.}

\REF\FM{O. Foda and T. Miwa, {\sl Int. Journ. Mod. Phys.}{\bf A7~1A}(1992)
  279.}

\REF\RIMS{B. Davies, O. Foda, M. Jimbo, T. Miwa and A. Nakayashiki,
  {\sl Commun. Math. Phys.}{\bf 151}(1993) 89.}

\REF\Y{ C. N. Yang, {\sl Phys. Rev. Lett.} {\bf 19}(1967)1312 ;
  {\sl Phys. Rev}{\bf 168}(1968)1920.  }

\REF\BAX{R. J. Baxter, {\sl Ann. Phys.} {\bf 58}(1977)1395;
  {\sl Exactly Solved Models in Statistical Mechanics}, Academic
  Press, New York  (1982).}

\REF\FAD{ L.H.  Takhtajan and L. D. Faddeev, {\sl Russ. Math. Survey}
  {\bf 34}(5) (1979) 11.}

\REF\BAZ{V. V. Bazhanov and N. Yu. Reshetikhin, {\sl Int. Journ. Mod.
  Phys.} {\bf 4}(1989) 115.}

\REF\SS{B. Sutherland and B. Shastry, {\sl Phys. Rev. Lett.} {\bf 65} (1990)
1833.}

\REF\KW{V. E. Korepin and A. C. T. Wu, {\sl Int. J. Mod. Phys.} {\bf B5} (1991)
497.}

\REF\YF{N. Yu and M. Fowler, {\sl Phys. Rev.} {\bf B46} (1992) 14583.}

\REF\B{M. V. Berry, {\sl Proc. R. Soc. London} {\bf A392} (1984) 45; see also,
{\sl Geometric Phases in Physics}, edited by A. Shapere and F. Wilczek
(World Scientific, Singapore, 1989).}

\REF\FT{L.D. Faddeev  and  L.H.  Takhtajan,  {\sl Phys. Lett.}{\bf 85A}
   (1981) 375.}

\REF\ABBAN{F. C. Alcaraz, M. Barber and M. Batchelor, {\sl Ann. of Phys.}
 {\bf 182}(1988) 280.}


\endpage

\chapter {Introduction}

 The one-dimensional spin one-half  XXZ chain with the Hamiltonian
$$
H=-{1\over 2}\sum^{N}_{j=1}(\sigma^{x}_{j}\sigma^{x}_{j+1}+
\sigma^{y}_{j}\sigma^{y}_{j+1}+\Delta\sigma^{z}_{j}\sigma^{z}_{j+1}) \eqn\Hxxz
$$
provides a soluble many-body problem  which  has been  intensively studied
   for more than two decades
  and is the subject of the present paper. \foot{  We find it impossible to
 exhaust all references on this subject.
 The partial list includes [\BETHE]-[\RIMS].    The developments in recent
 years
    include the relation to the  study of  finite size corrections  as well as
  the approach based on the affine $U_{q}(  \hat{sl} (2)),$
   which  are seen, for example, in [\ABB], [\PT], [\FM], [\RIMS]. }
 Here $\sigma^{x,y,z}_{j}$ are the Pauli
   matrices at site $j$ and $\Delta$ denotes the parameter of  anisotropy.
  Let a generic M-body state  of the  model be represented by
$$
|\chi\bigr> = \sum_{x_1=1}^N\cdots\sum_{x_M=1}^N\chi(x_1,\ldots,x_M)
              \prod_{j=1}^M\sigma_{x_j}^{-}|\uparrow\bigr>,
\eqn\Bethest
$$
where $|\uparrow\bigr>$ is the state with all spins up
 and $\sigma_x^{-}$ is the spin-lowering matrix  at site $x$.
The eigenfunction $\chi(x_1,\ldots,x_M)$  is of the form proposed by
 Bethe [\BETHE]:
$$
\eqalign{
& \chi(x_1,\ldots,x_M|p_1,\ldots,p_M) \cr
= &
\left[
  \prod_{M\geq b>a\geq 1}
  \epsilon(x_b-x_a)
\right]
\sum_Q(-1)^{|Q|}{\rm exp}
\left[
  i\sum_{a=1}^Mx_ap_{Qa}+
  {i\over 2}\sum_{M\geq b>a\geq 1}\theta(p_{Qb},p_{Qa})\epsilon(x_b-x_a)
\right]
},
\eqn\Bethewf
$$
where $\epsilon(x)$ is the sign function, the summation is with respect to
permutations $Q$ of the momenta $p$ and $\theta$ is the two-particle
phase shift.
For two particles with momenta $p$ and $q$, the explicit form of the phase
shift is
$\theta(p,q) = 2 {\rm arctan}\left[  { {\Delta{\rm sin}[(p-q)/2]}\over
    {{\rm cos}[(p+q)/2] - \Delta{\rm cos}[(p-q)/2] } }\right].$
The energy of this eigenstate is found to be
$E = -{N\over 2}\Delta + 2\sum_{j=1}^M(\Delta-{\rm cos}p_j).$
The second term is the sum of the single-particle energies
$\varepsilon_j = 2(\Delta-{\rm cos}p_j)$.
The operator
$ M = \sum_{j=1}^N {1 \over 2} (1 - \sigma_j^z) $ is conserved  and
 the case of half-filling  corresponds to $M = N/2$.
The complete reducibility of the problem to the two-body problem with
  the dynamical phase shift as well as the rigorous notion of the
  quasiparticle excitations  is the hallmark of the integrable
  quantum systems whose mathematical structure is captured by
  the Yang-Baxter relation [\Y] [\BAX] and the attendant quantum
 inverse scattering  formalism [\FAD].

The boundary condition is a heart to this problem as this
 determines the spectra of the model and is  referred to as
  the Bethe ansatz equation in this paper.
Let us now impose, contrary to the standard periodic boundary
  condition,  the  twisted boundary condition
$$
\chi(x_1,\ldots,x_j+N,\ldots,x_M) = {\rm e}^{i\Phi}
\chi(x_1,\ldots,x_j  ,\ldots,x_M),
\eqn\eq
$$
  where  $\Phi$ is a twist angle.\foot{ The twist angle plays an important
  role in projecting out some states to construct another model.
   See [\PS] [\BAZ]  for example.}
This leads to the system of transcendental equations for the set of momenta
$\{p_j\}$
$$
Np_j + \sum_{k=1(\neq j)}^M\theta(p_j,p_k) = 2\pi I_j + \Phi,
\eqn\trans
$$
where $j=1,\ldots,M$.  We denote by
$\{I_j\}$  the set of half-integers for even $M$ or integers for odd $M$.

  In this paper, we study ground state properties of the system
   when  the twist angle $\Phi$ is varied. The change of  an
  external parameter like $\Phi$ in integrable systems will correspond
  to an adiabatic change in  more general systems  which do not possess
  an infinite number of conservation laws .  The process we consider
  may be called adiabatic process in this sense.
  The major purpose  of this paper is to examine  the motions of the roots of
 eq. \trans under this process  and to determine the momenta and the phase
 shift functions for the final configuration, given  the ones
 for the initial configuration.

It is well-known that the system can be represented by  the Jordan-Wigner
  fermions.
 The state with spin-up is regarded   as an unoccupied state and the
  state with spin-down  as an occupied state.
The annihilation operator $C_j$ and the creation operator $C_j^\dagger$
are respectively defined by
$C_j = {1\over 2}(\sigma_j^x + i\sigma_j^y) \prod_{k<j} \sigma_k^z,
C_j^\dagger = {1\over 2}(\sigma_j^x - i\sigma_j^y) \prod_{k<j} \sigma_k^z,$
which satisfy anticommutation relations
$\{ C_j, C_k^\dagger \} = \delta_{j,k}.$
The Hamiltonian \Hxxz\ is rewritten in terms of lattice
fermions as
$$
H = -\sum_j (C_j^\dagger C_{j+1} + C_{j+1}^\dagger C_j)
    -2\Delta\sum_j (C_j^\dagger C_j - {1\over 2})
     (C_{j+1}^\dagger C_{j+1} - {1\over 2}).
\eqn\Hferm
$$
The twisted boundary condition  can be attributed to the twisting of the
  operators $ C_j = {\rm e}^{i\Phi}C_{j+N},\qquad
C_{j}^\dagger = {\rm e}^{-i\Phi} C_{j+N}^\dagger.$
  The conserved operator  $M=\sum_{j=1}^N C_j C_j^{\dagger}$ measures
 the fermion number.
The twisted boundary condition is related to the Aharonov-Bohm effect
 if we consider the $M$ fermions carrying a charge $(-q)$ on the ring
 threaded by the magnetic flux $\Phi/q$: as each fermion  goes around the ring,
 it picks up the Aharonov-Bohm phase $\Phi$.

  The change of  the twist angle $\Phi$
 is equivalent to varying the magnetic flux penetrating the ring which
generates
an electric field around the ring.
The whole spectrum of Hamiltonian is periodic with respect to $\Phi$
with  period $2\pi$.
If we follow each individual energy level, this is not necessarily  the
 case  however.
The period of  the ground state, in particular, is a nontrivial quantity
 to be determined.
In the case  of no interaction  $\Delta =0$,  the period can be  obtained
  by a classical  consideration alone.
 All negative energy pseudoparticles are accelerated in the electric field $E$
  generated by the change of the flux $\Delta \Phi$  through
$\Delta \Phi = \Phi(t_I) - \Phi(t_F) = \int_{t_F}^{t_I} dt \oint dx qE.$
The change of the momentum is
$\Delta p_j = \int_{t_I}^{t_F} dt qE = {\Delta \Phi \over N}$
where  $N= \oint dx$.
Thus the period of the ground state is $\Delta \Phi = 2N\pi$.
In this process all negative energy pseudoparticle move from the first
Brillouin zone $-\pi < p < \pi$ to the second Brillouin zone $\pi < p < 3\pi$.

  A completely different picture is known to be obtained as soon as  the
  the interaction is turned on [\SS].  In [\SS],   the regime
  $ -1 < \Delta < 0 $ is considered.  It is found that
  only one negative energy pseudoparticle   moves from
 the first  Brillouin zone $-\pi < p < \pi$ to the
 second Brillouin zone $\pi < p < 3\pi$.  The period of the ground
 state is found to be  $\Delta \Phi = 4\pi$.
During this process a level crossing occurs.
The ground state becomes  the excited state for the system
  with the original periodic boundary condition
  at  $\Phi =2\pi$ and comes back to the original ground  state at $\Phi =
4\pi$.
The Berry phase attendant with this process has been considered
 in ref. [\KW] and is found to be $\pi$.

In the regime $0<\Delta<1$, more interesting
events take place.
It is pointed out in ref. [\YF]  that,
  in the region $\cos(\pi/n+1)>\Delta>\cos(\pi/n)$,
 a bound state  of the $n$-string is formed
during the process.
Only these roots of the Bethe ansatz equation forming the $n$-string
 move from the first Brillouin zone $-\pi < p < \pi$ to
 the second Brillouin zone $\pi < p < 3\pi$. The period of the ground state
 is   again found to be $\Delta \Phi = 4 \pi$.

The thrusts of the present paper  consist of the following several points.
 In section $2$, we   consider the region $0< \Delta < 1$.  In the segment
   $\cos(\pi/n+1)>\Delta>\cos(\pi/n)$  for $ n= 2,3,\cdots$,
 we determine  the motions of
  the roots of the Bethe ansatz eq. \trans  in the rapidity
 plane \foot{In this paper, the words ``root'' and ``rapidity'' are
  used interchangeably.}.
  This is done on the basis of our numerical and analytic study at a finite
 size system.
  Here we find  intriguing motions of  the $n$-string during the
  process.
   In order to state our results here  more clearly, let us recall  that
the $n$-string is a group of rapidities which differ in their imaginary
parts by $2i(\pi-\mu)$ and center on the $i\pi$ line
(Im$\lambda=\pi$):
$$
\eqalign{\lambda_{k}^{(n)} &= [\lambda^{(n)}+ i\pi] + i(\pi-\mu)(n+1-2k)
           + \delta_k^{(n)} \cr
          & \qquad\qquad k=1,\ldots,n\,}
\eqn\nstr
$$
where $[\lambda^{(n)} + i\pi]$ is the center of the $n$-string and
$\delta_k^{(n)}$ is a deviation due to the finiteness of the  system from
 the string found in the infinite volume limit. \foot{ In this paper, we deal
with
  a natural formation of the $n$-string configuration.  The Takahashi condition
[\T]  will be automatically satisfied.}
This deviation is at most of order $1/N$.
We  find, however, that it is this deviation  which plays  a  vital role in
   determining the global structure of
  the motion of the roots under the  process.
  To be more explicit, we find  that, in the case of  the  two-string,
   the roots collide
  into  the branch points  of the momenta and
  those of the phase shift functions
  perpendicularly  to  the imaginary axis.   In the case of the three-string,
   we find that the roots fluctuate around the center of the string
 due to the finite size correction  $\delta_{k}^{(n)}$
 mentioned above.   The case of the $n$-string  is found to a generalization
  of these: for $n$ odd, the roots oscillate around the center of the
  string while, for $n$ even, in addition to the
  fluctuation,  the two  roots  in the middle collide into the branch points.
This fluctuation of string state creates a winding
 around   the cut of the momentum and the phase shift function
  and determines the final configurations of these.
 These are presented in section $2$.

  In section $3$, we consider   the regime  $ \Delta \leq -1$
   and determine the motion of the roots.
   From this we determine the period,  the final configurations of the momenta
 and the phase shift functions  and the  Berry phase.
   These values are found to be the same as those derived in ref.[\SS], [\KW]
  in $-1 < \Delta< 0$. We, therefore, establish  the continuity
  at $\Delta =-1$.

  In section $4$,
   We compute the Berry phase in the regime  $ 0 < \Delta < 1 $  and
 consider the physics associated with it.
  By piecing through the computation done in [\KW] and ours, we
  are led to  propose that the Berry phase provides a measure of the statistics
 of  the quasiparticle (or the bound state) involved  in the process.
 We consider  statistics of particles as  the Berry phase
  arising through $p$-space monodromy.

   In Appendix A,  we summarize our analytic study of the finite size system
  at  $M=2, N=4$.   In  Appendix B and Appendix C, we
 present  proofs of technical lemmas
  necessary in  $\S 2-1$ and $\S 4$  respectively.

\endpage

\chapter{Motions of String Solutions under a Varying Twist Angle
  in the Regime   $ 0< \Delta < 1$}

  In the regime
$-1<\Delta<1$, the momentum $p$, the phase shift $\theta$ and the total energy
  $E$ are
parametrized by rapidities ${\lambda}_j $;
$$
p(\lambda) = -i{\rm ln}\
\left[
  -{{{\rm sinh}{1\over 2}(\lambda-i\mu)}\over
    {{\rm sinh}{1\over 2}(\lambda+i\mu)}}
\right],
\eqn\moment
$$
$$
\theta(\lambda_1,\lambda_2) = i{\rm ln}\
\left[
  -{{{\rm sinh}{1\over 2}(\lambda_1-\lambda_2-2i\mu)}\over
    {{\rm sinh}{1\over 2}(\lambda_1-\lambda_2+2i\mu)}}
\right],
\eqn\phashiftr
$$
$$
E = {N \over 2} {\rm cos} \mu + \sum_{j=1}^M { {-2 {\rm sin}^2 \mu} \over
{{\rm cosh} \lambda_j - {\rm cos} \mu} },
\eqn\eq
$$
where $\Delta=-{\rm cos}\,\mu$.  Let us imagine a closed circuit in the
  $\lambda$ plane consisting of the real axis  and the $i\pi$ line connected
 at plus and minus infinity.
The momentum $p(\lambda)$ increases monotonically as a function of
$\lambda$ from $-(\pi-\mu)$ at $\lambda=-\infty$ to $(\pi-\mu)$
at $\lambda=\infty$.  ( See Figure 2.1).
If $p$  increases further, $\lambda$ moves backward along
the $i\pi$ line from $+\infty + i\pi$.
The momentum $p(\lambda)$ varies from $(\pi-\mu)$ at
$\lambda=\infty+i\pi$ to $2\pi-(\pi-\mu)$ at $\lambda= -\infty +i\pi$.
When the rapidity goes around this closed circuit, the momentum $p$
increases by $2\pi$, since the path encircles a logarithmic branch
cut as a function of $\lambda$.
We note that the single-particle energy
$$
\varepsilon_j = { {-2 {\rm sin}^2 \mu} \over
{{\rm cosh} \lambda_j - {\rm cos} \mu} }
\eqn\eq
$$
is negative  for ${\rm Im}\,\lambda_j=0$ and positive for
${\rm Im}\,\lambda_j=\pi$.

The behavior of the phase shift function $\theta(\lambda)$
depends on the sign of $\Delta$. ( See Figure 2.1.)
When $\Delta<0\ (0<\mu<{\pi\over 2})$, the phase shift
$\theta(\lambda)$ decreases from $(\pi-2\mu)$ at $\lambda=-\infty$
to $-(\pi-2\mu)$ at $\lambda=\infty$.
The phase shift further decreases from $-(\pi-2\mu)$ at $\lambda=\infty+i\pi$
to
$-2\pi+(\pi-2\mu)$ at $\lambda=-\infty+i\pi$.
As the relative rapidity goes around the circuit, the phase shift
{\it decreases} by $2\pi$.
When $\Delta>0\ ({\pi\over 2}<\mu< \pi)$, the phase shift $\theta(\lambda)$
increases from $-(2\mu-\pi)$ at $\lambda=-\infty$ to $(2\mu-\pi)$ at
$\lambda=\infty$.
The phase shift further increases from $(2\mu-\pi)$ at $\lambda=\infty+i\pi$ to
$2\pi-(2\mu-\pi)$ at $\lambda=-\infty+i\pi$.
As the relative rapidity goes around this circuit, the phase shift
{\it increases} by $2\pi$.
This difference plays a central role
 in determining the behavior of the rapidities as seen later.

\midinsert
\epsfysize=15cm
$$\epsffile{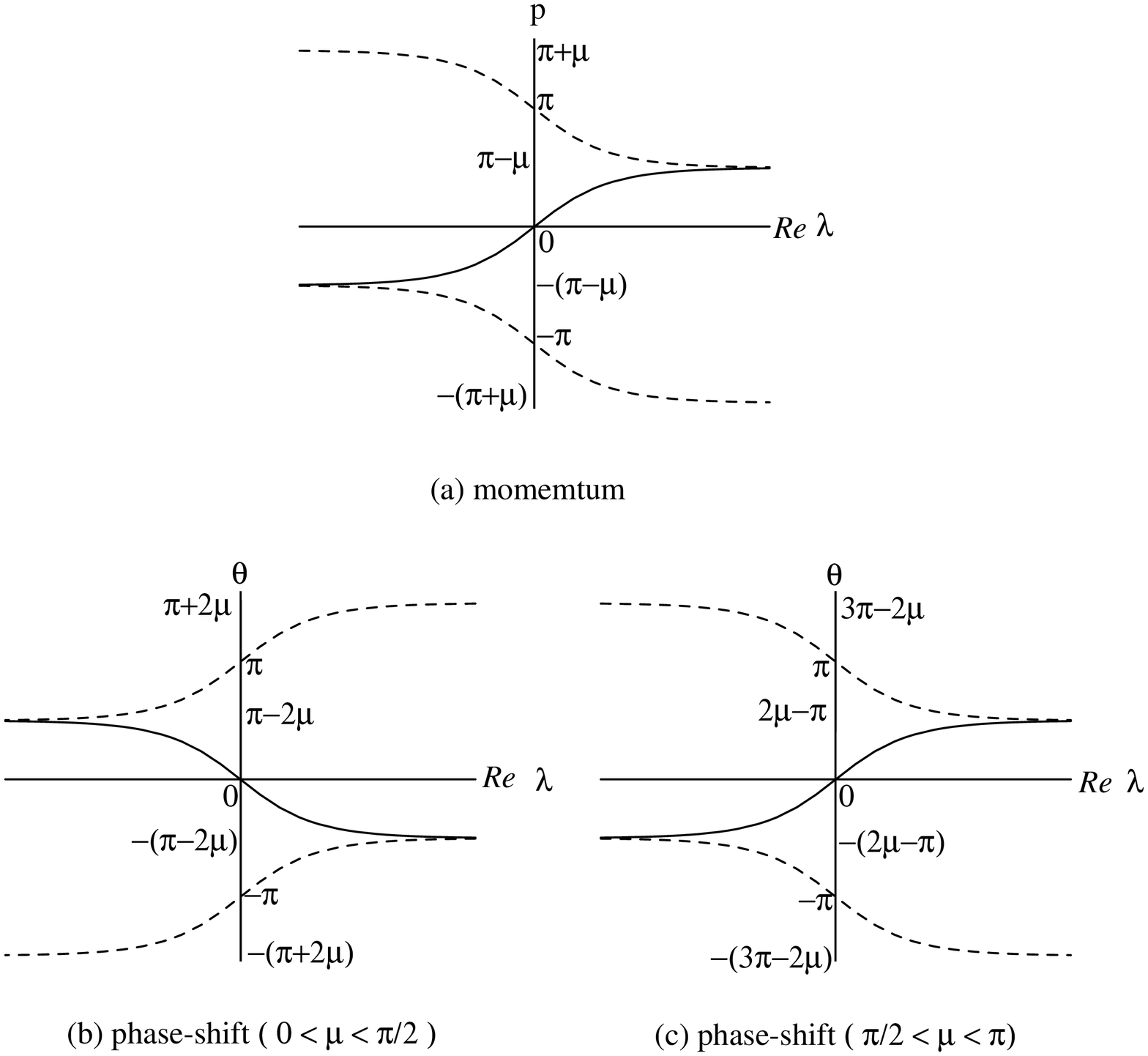}$$
\nobreak
\narrower
\singlespace
\noindent
Figure 2-1. The momentum  function
 and  the phase-shift function for $Im \lambda = 0$
 (normal lines) and for $Im \lambda = \pi$ (dotted lines).
\medskip
\endinsert

Before we study the region $ 0<\Delta < 1 $, we briefly look at the
region $-1<\Delta<0$ considered in ref.[\SS].
All the rapidities are symmetrically arranged on the real axis
at $\Phi=0$.
We denote them  by
 $-\infty <\lambda_1<\lambda_2<\cdots<\lambda_M< \infty$.
The largest rapidity $\lambda_M$ goes to infinity at $\Phi=2(\pi-\mu)$.
Here the $\lambda_M$ jumps from the real axis where a negative-energy mode lies
to the $i \pi$ line where a positive-energy mode lies.
This means a particle creation.
When $\Phi$ increases from $2(\pi-\mu)$, $\lambda_M$ moves backward along
the $i\pi$ line and the other rapidities remain on the real axis.
The $\lambda_M$ on $i\pi$ line  may be called one-string.
When $\Phi=2\pi$, the rapidity $\lambda_M$ reaches $i\pi$.
When $\Phi=2\pi-2(\pi-\mu)$, $\lambda_M$ goes to $-\infty+i\pi$ and
jumps onto the real axis.
This means a particle annihilation.
The state comes back to the original ground state at $\Phi=4\pi$.
The relation between the initial and the final momentum and the one
 between the initial and the final phase shift are the same as
   in the regime $ \Delta < -1$, which we will show
  in \S 4.
The Berry phase of the ground state wave function
  in this regime has been calculated
  in ref.[\KW] to be $\pi$.

In the case of half-filling  $M=N/2$  which we consider, the picture
 of the Dirac sea emerges as follows.
The eq. \trans\ reduces to $p_j = (I_j / N)\times 2\pi$ with the condition
 $\Phi = 0$.
The single-particle energy  is
$\varepsilon_j=-2{\rm cos}p_j = -2{\rm cos}(2 \pi I_j / N)$.
The negative energy modes  are in the region $|I_j|  \leq N/4$   and
 are given by the set
$$
I_j = -{M-1\over 2}, -{M-3\over 2},\ldots,{M-1\over 2} .
\eqn\Betheqn
$$
The same set $\{I_j\}$ turns out to provide
 the ground state of the half-filling sector in the interacting case.

In the next two subsections, we consider the case in which only a two-string
 is formed ($ 0< \Delta < \cos ( \pi/3) $)   and
 the case   in which a three-string is formed
   ($ \cos ( \pi/3) < \Delta  <  \cos ( \pi/4) $).
 The conclusion we will draw in the former case is
 mainly based on the analytic study  we have made
 at a finite size system $M=2, N=4$,
  which we present in  Appendix A.
  The conclusion we will draw in the latter case is   originally derived from
 the  numerical  study we  have carried out  in the case $M=3, N=6$.
   In both cases, we check the consistency condition that the momenta
 and the phase shift functions for the final configuration
 follow from those for the initial one by using
 the relations we derive from  the global behavior of the rapidities.
   We find  that this consistency condition is  a nontrivial check to make.
   In the last subsection,  we are able to  see that the global behavior of
   the rapidities  we find in the case of $n$-string formation
   ( $ \cos (\pi/n)  < \Delta < \cos (\pi/n+1)$ )  passes  this consistency
  check for any $n$.   This check is  very stringent and
 consolidates our conclusion.

  It will be helpful to point out here that one can predict a few things about
   the behavior of the rapidities
   without looking at the details of  eq. \trans.
 As we vary the twist angle continuously,
  an individual momentum, which is a root of  eq. \trans,
  must vary continuously.
   The only place in which
 a set of rapidities can jump to form a string   while
  keeping the momenta continuous is the point at infinity  ( i.e. the
 real part of the rapidity is infinite.  See eq. \moment ).
  This is also the place for the
 formation of a zero-energy bound state. ( See eq. \eq .)
 The value of the twist angle    where the formation of $n$-string at infinity
  can occur can be obtained from  eqs. \trans , \moment .
  We start from  eq. \trans    for $M$ rapidities
  and send the largest ( or smallest)
 $n$ rapidities to plus ( or minus ) infinity, which subsequently form an
  $n$-string. Only when the twist angle $ \Phi = + ( or -) n (\pi - \mu)$,
  the rest of the $M-n$ rapidities are kept symmetric with respect
  to the origin, staying in the  ground state configuration.
  In other words, only at this value  the ground state configuration
  for $M$ rapidities reduces to that for the  $M-n$ rapidities.

\section{ Ground state in  the case of two string}

We now study the more specific case $0<\Delta<{\rm cos}(\pi/3)$\
$({\textstyle \pi\over 2}<\mu<{\textstyle 2\over 3}\pi)$ where the two-string
is expected to be formed  during the process.
 For convenience  we describe the  process to be associated with the change of
$\Phi$ from $-2\pi$ to $2\pi$.
We first examine the latter half $0\leq\Phi<2\pi$ of the process.
When $\Phi$ increases from zero the largest rapidity
$\lambda_M$ reaches $\infty$ at $\Phi=2(\pi-\mu)$ just as in the
  $ -1 < \Delta < 0 $ case.
The rapidity $\lambda_M$ jumps to the $i\pi$ line and begins to move
backward along the $i \pi $ line from $\infty+i\pi$ .
The dramatic change  arises when $\Phi$ increases further.
The rapidity $\lambda_M$ on the $i\pi$ line returns to $\infty+i\pi$.
The second largest rapidity $\lambda_{M-1}$ on the real axis goes to
infinity.
The two rapidities $\lambda_M$ and $\lambda_{M-1}$ attract each
other.
At $\Phi=4(\pi-\mu)$, the two rapidities reach
$\lambda_M=\infty+i\pi$ and $\lambda_{M-1}=\infty$.
The two largest rapidities then
jump together by $i\pi/2$ to the imaginary direction and
they form a two-string:
$\lambda_M    =\infty+{\textstyle 3\over 2}i\pi$ and
$\lambda_{M-1}=\infty+{\textstyle 1\over 2}i\pi$.
This means  a formation of the bound state  over the vacuum.
The two rapidities are separated by $i\pi$ and their center is
on the $i\pi$ line.
So the deviation due to the finite size in \nstr\ is
  given by $\delta_M=\delta_1^{(2)}=i\mu$ and
$\delta_{M-1}=\delta_2^{(2)}=-i\mu$.

When we further increase  $\Phi$, the two-string moves backward from
plus infinity.
The center moves on the $i\pi$ line from $\infty+i\pi$.
In this process the difference between $\lambda_M$ and $\lambda_{M-1}$
decreases from
$i\pi$.
At $\Phi=2\pi$, the two-string arrives at the imaginary axis:
$$
\eqalign{\lambda_M     &= i(2\pi-\mu),\cr
         \lambda_{M-1} &= i\mu}
\eqn\IIstr
$$
We note that the relative rapidity $(\lambda_M - \lambda_{M-1})$ is
 $2i(\pi-\mu)$, and the deviations
$\delta_1^{(2)}$ and $\delta_2^{(2)}$ in \nstr\ vanishes.
It is interesting that the momenta $p(\lambda_M)$,
$p(\lambda_{M-1})$ and the phase shift $\theta(\lambda_M,\lambda_{M-1})$
are divergent at this point.
The points \IIstr\ are  the branch points of both
 the momentum $p(\lambda)$
and the phase shift $\theta(\lambda)$, which are logarithmic
functions \moment\ and \phashiftr\ respectively.
If the rapidity $\lambda_j$ goes around the branch point
$\lambda=i\mu\ (\lambda=i(2\pi-\mu))$ counterclockwise,
the momentum $p(\lambda_j)$ increases (decreases) by $2\pi$.
If the relative rapidity $\lambda$ goes around the branch point
$\lambda=2i(\pi-\mu)$ counterclockwise, the phase shift increases by
$2\pi$.
Therefore the behavior of the rapidities around these branch points
  has a global meaning and is necessary to determine
 the final configuration of the system.
The paths of the rapidities $\lambda_M$, $\lambda_{M-1}$ and
  those of the relative rapidity $\lambda_M - \lambda_{M-1}$
around the branch points are plotted in Fig. 2-2 and Fig. 2-3 respectively.
The $\lambda_M$ and $\lambda_{M-1}$ collide into the singularities
perpendicularly to the imaginary axis (Fig. 2-2).
The value of the relative rapidity $\lambda_M-\lambda_{M-1}$ reduces from
$i\pi$ to $2i(\pi-\mu)$ (Fig. 2-3).
This behavior of the rapidities is checked both by the numerical
 calculation and by the analytic calculation in the case of $M=2$ which
 is described  in Appendix A.
The first half process in $-2\pi<\Phi\leq 0$ is similar to the latter
 half process.
At $ \Phi = -2 \pi $, the two-string starts from the singular points
$ \lambda_1 = i(2\pi-\mu), \lambda_2 = i\mu $.
When $\Phi=-4(\pi-\mu)$ the two-string reaches
$\lambda_1=-\infty+i{\textstyle 3\over 2}\pi$ and
$\lambda_2=-\infty+i{\textstyle \pi\over 2}$.
After jumping by $-i\pi/2$, $\lambda_1$ and $\lambda_2$ move on the
$i\pi$ line and the real axis respectively.
Finally $\lambda_1$ jumps onto the real axis at $\Phi=-2(\pi-\mu)$.

\midinsert
\epsfysize=10cm
$$\epsffile{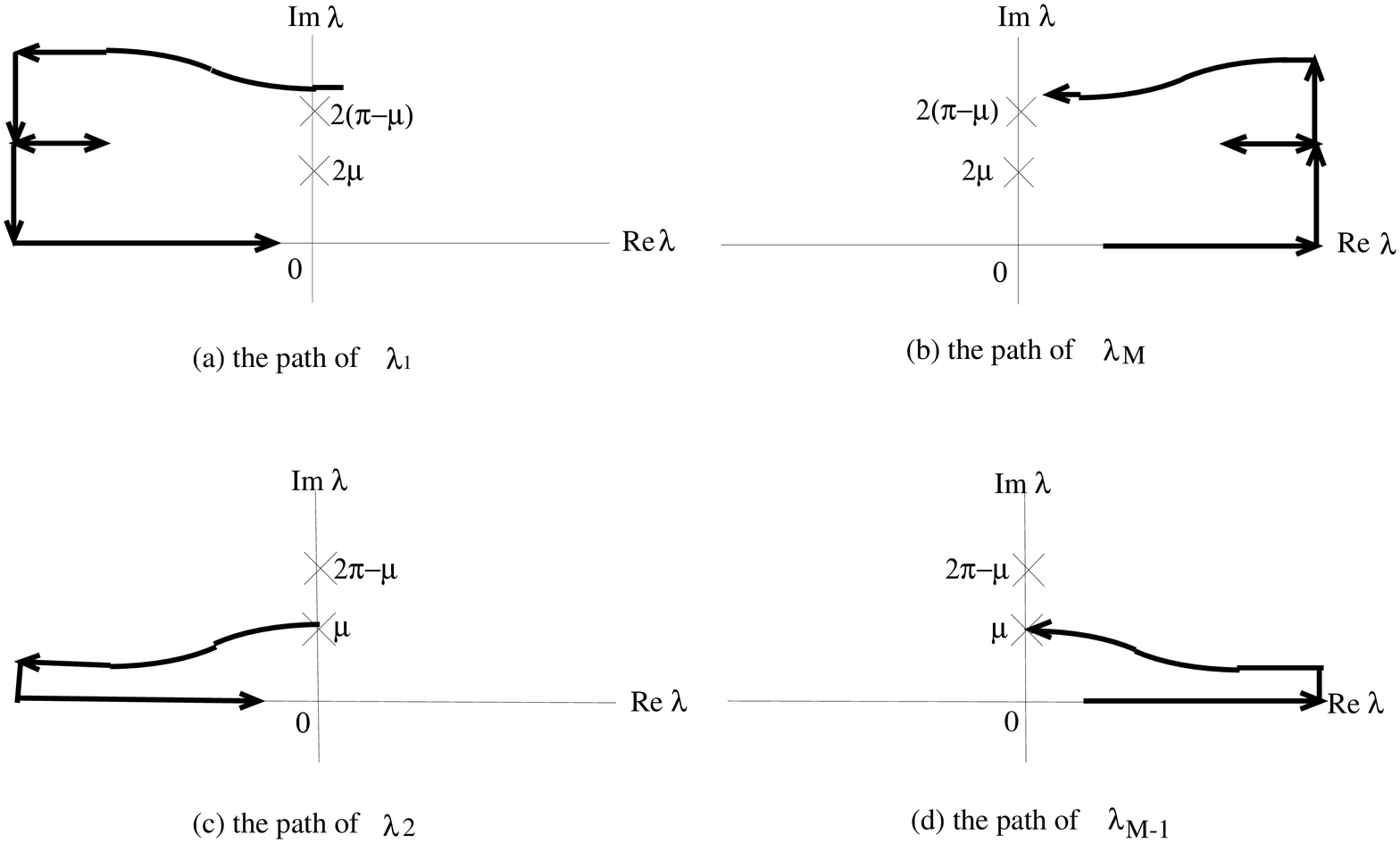}$$
\nobreak
\narrower
\singlespace
\noindent
Figure 2-2. The paths of the rapidities in case of $\pi / 2 < \mu < 2\pi / 3$.
\medskip
\endinsert

\midinsert
\epsfysize=14.5cm
$$\epsffile{pathb.eps}$$
\nobreak
\narrower
\singlespace
\noindent
Figure 2-3. The paths of the relative rapidities in the case $\pi / 2 < \mu <
2\pi / 3$.
\medskip
\endinsert

The relations between the initial state $(\Phi=-2\pi)$ and
the final state $(\Phi=2\pi)$ are determined from this behavior of the
rapidities.
They are rearranged as
$$
\eqalign{\lambda_M^{(f)}     &= \lambda_1^{(i)},   \qquad
         \lambda_{M-1}^{(f)}  = \lambda_2^{(i)},   \cr
         \lambda_j^{(f)}     &= \lambda_{j+2}^{(i)}}\ \ (j=1, \cdots , M-2).
\eqn\IIpe
$$
Let us determine the relation between
$\{ p_j^{(i)}, \theta(p_j^{(i)},p_k^{(i)})\}$ and
$\{ p_j^{(f)}, \theta(p_j^{(f)},p_k^{(f)})\}$.
The cut of the function $p(\lambda)$ extends from the branch point $i\mu$ to
another branch point $i(2\pi - \mu)$.
The path of $\lambda_1$ starts from the branch point $i(2\pi -\mu)$ and
 that of $\lambda_M$
reaches the same branch point $i(2\pi -\mu)$  (Fig. 2-2).
To determine the relation between $p_{M}^{(f)}$ and $p_1^{(i)}$, we follow a
circuit consisting of the path of $\lambda_M$, that of $\lambda_1$ and the real
axis.
We start from the branch point $\lambda_1^{(i)} = i(2\pi - \mu)$
 perpendicularly to the imaginary axis.
Finally we come to the same branch point $\lambda_M^{(f)} = i(2\pi - \mu)$
perpendicularly to the imaginary axis after running through the circuit.
 The circuit encircles the 'half' of the branch point $\lambda=i(2\pi -\mu)$ of
the function $p(\lambda )$.
We conclude that $p(\lambda_M^{(f)})$ is larger than $p(\lambda_1^{(i)})$ by
$\pi$:
$$
p_M^{(f)}=p_1^{(i)}+\pi.
\eqn\IIpa
$$
Similarly we find the relation
$$
p_{M-1}^{(f)}=p_2^{(i)}+\pi.
\eqn\IIpb
$$
The other relations are trivial
$$
p_j^{(f)}=p_{j+2}^{(f)}\qquad (j=1,\ldots,M-2).
\eqn\IIpc
$$
The cut of the function $\theta(\lambda)$ extends from the branch point
$2i(\pi-\mu)$
to another branch point $2i\mu$.
To determine the relation between $\theta(p_M^{(f)},p_{M-1}^{(f)})$ and
$\theta(p_1^{(i)},p_2^{(i)})$ we consider the circuit consisting of the paths
of the relative rapidities $\lambda_M-\lambda_{M-1}$, $\lambda_1-\lambda_2$ and
 the real axis.
This circuit crosses this cut (Fig. 2-3).
Therefore the relation between the two phase shifts is
$$
\theta(p_M^{(f)},p_{M-1}^{(f)}) = \theta(p_1^{(i)},p_2^{(i)})+2\pi.
\eqn\eq
$$
To consider the other relations for the phase shifts,
recall that the difference in the two rapidities in the two-string is larger
 than $2(\pi-\mu)$ in this process. From this
 we say the relation ${\rm Im}\,(\lambda_M)\geq \pi + ( \pi - \mu) = 2\pi-\mu$
and ${\rm Im}\,(\lambda_{M-1})\leq \pi -(\pi - \mu) = \mu$ in the two-string.
Therefore
${\rm Im}\,(\lambda_M    -\lambda_j)\geq 2\pi-\mu>2\mu\  (j\neq M,M-1)$
and
${\rm Im}\,(\lambda_{M-1}-\lambda_j)\leq \mu<2(\pi-\mu)\ (j\neq M,M-1)$
are satisfied in the region of our
consideration ${\textstyle \pi\over 2}<\mu<{\textstyle 3\over 2}\pi$.
This means that the circuit consisting of the paths of $\lambda_M-\lambda_j$,
 that of $\lambda_{1}-\lambda_{j+2}$ and the real axis
 does not cross the cut of the phase shift function (Fig. 2-3).
Similarly the circuit consisting of the paths of $\lambda_{M-1}-\lambda_{j}$,
$\lambda_2-\lambda_{j+2}$ and the real axis does not cross the cut (Fig. 2-3).
 From these results we obtain the relations
$$
\eqalign{
\theta(p_M^{(f)}    ,p_j^{(f)}) &= \theta(p_1^{(i)},p_{j+2}^{(i)}),\cr
\theta(p_{M-1}^{(f)},p_j^{(f)}) &= \theta(p_2^{(i)},p_{j+2}^{(i)}),}
\eqn\eq
$$
where $j=1,\ldots,M-2$.
The other relations are trivial
$$
\theta(p_j^{(f)},p_k^{(f)}) = \theta(p_{j+2}^{(i)},p_{k+2}^{(i)}),
\eqn\IIpd
$$
where $j,k=1,\ldots,M-2$.

The set $\{p_j^{(i)}\}$ and the set $\{p_j^{(f)}\}$ are not equivalent
even up to $2\pi$, as seen in \IIpa\ and \IIpb.
In spite of this, the initial state $(\Phi=-2\pi)$ and the final state
$(\Phi=2\pi)$ are  found to be the same.  We show this in Appendix B.

Our conclusion \IIpa, \IIpb\ and \IIpc\ are different from the one
  derived in [\YF].
In ref. [\YF], it was concluded  that the two-string never comes close enough
to the $i\pi$ line.
They have obtained the relations $p_M^{(f)}=p_1^{(i)}$,
$p_{M-1}^{(f)}=p_2^{(i)}$ and
$p_j^{(f)}=p_{j+2}^{(i)}$.
 Their conclusion is not valid however.
By summing over  the relations \trans\ for all $j\ (j=1,\ldots,M)$,
we obtain a simple relation
$$
\sum_{j=1}^Mp_j={\Phi\over 2},
\eqn\eq
$$
which must be satisfied.
Their conclusion does not satisfy this simple consistency condition.

Finally, let us check briefly the consistency between the boundary conditions
\trans and our relations \IIpa --\IIpd .
The boundary conditions for the final state are
$$
2Mp_j^{(f)}+\sum_{k\neq j}\theta(p_j^{(f)},p_k^{(f)})
= 2\pi\left({2j-M-1\over 2}\right) + 2\pi, \ \
 \quad (j=1,\ldots,M).
\eqn\eq
$$
These equations are rewritten in terms of the initial set by using
\IIpa --\IIpd .
$$
2M(p_M^i+\pi)+\left(\sum_{k\neq 1}\theta(p_1^{(i)},p_k^{(i)})+2\pi\right)
= 2\pi\left({M-1\over 2}\right) + 2\pi,
\eqn\eq
$$
$$
2M(p_2^{(i)}+\pi)+\left(\sum_{k\neq 2}\theta(p_2^{(i)},p_k^{(i)})-2\pi\right)
= 2\pi\left({M-3\over 2}\right) + 2\pi,
\eqn\eq
$$
$$
\eqalign{
2Mp_{j+2}^{(i)}+\sum_{k\neq j+2}\theta(p_{j+2}^{(i)},p_k^{(i)})
&= 2\pi\left({2j-M-1\over 2}\right) + 2\pi, \cr
& \quad (j=1,\ldots,M-2).}
\eqn\eq
$$
They are equivalent to the boundary conditions for the initial state
$$
\eqalign{
2Mp_{j+2}^{(i)}+\sum_{k\neq j+2}\theta(p_{j+2}^{(i)},p_k^{(i)})
&= 2\pi\left({2(j+2)-M-1\over 2}\right) - 2\pi, \cr
& \quad (j=  -1, 0,\ldots,M-2).}
\eqn\eq
$$

%
\section{Ground state in the case of three string}

Let us now turn to  the case of three-string which is formed in the region
${\rm cos}\,(\pi/3)< \Delta <{\rm cos}\,(\pi/4),\
 ({\textstyle 2\over 3}\pi<\mu<{\textstyle 3\over 4}\pi)$.
  We first describe the behavior of the rapidities  we have found in this
region.
When $\Phi$ increases from zero to $2(\pi-\mu)$, the
largest rapidity $\lambda_M$ goes to infinity and jumps onto the
$i\pi$ line.
When $\Phi$ increases from $2(\pi-\mu)$ to
$4(\pi-\mu)$, the second largest rapidity $\lambda_{M-1}$
goes to infinity and the two rapidities jump by $i\pi/2$ simultaneously.
Thus the two rapidities $\lambda_M=\infty+{\textstyle 3\over 2}i\pi$
and $\lambda_{M-1}=\infty+{\textstyle 1\over 2}i\pi$ form a two-string
as seen in \S 2.1.
As $\Phi$ increases further from $4(\pi-\mu)$, the two-string starts
 moving from
plus infinity. The center of the two-string is on the $i\pi$ line and the
distance between $\lambda_M$ and $\lambda_{M-1}$ becomes shorter.
The two-string and the third largest rapidity $\lambda_{M-2}$ attract
 each other.
When $\Phi=6(\pi-\mu)$, the two-string and $\lambda_{M-2}$ go to infinity.
Here the distance between $\lambda_{M}$ and $\lambda_{M-1}$ becomes $2\pi/3$.
 Three rapidities are, therefore,
 $\lambda_M=\infty+{\textstyle 4\over 3}i\pi$,
\ $\lambda_{M-1}=\infty+{\textstyle 2\over 3}i\pi$,
\ $\lambda_{M-2}=\infty$.
Here the three rapidities jump by
${\textstyle\pi\over 3}i$.
They form a three-string where
$\lambda_M=\infty+{\textstyle 5\over 3}i\pi$,\
$\lambda_{M-1}=\infty+i\pi$ and
$\lambda_{M-2}=\infty+{\textstyle 1\over 3}i\pi$.
The  deviation of the length of the string  due to the finite size
 in \nstr is
$\delta_M\equiv\delta_1^{(3)}=2i({\textstyle 2\over 3}\pi-\mu)$,\
$\delta_{M-1}\equiv\delta_2^{(3)}=0$ and
$\delta_{M-2}\equiv\delta_3^{(3)}=-2i({\textstyle 2\over 3}\pi-\mu)$.

When $\Phi$ increases from $6(\pi-\mu)$ to $4\pi-6(\pi-\mu)$,
the center of the three-string moves from $\infty+i\pi$ to $-\infty+i\pi$.
The three-string fluctuates in this process (Fig. 2-4).
It is interesting to note that the deviation $\delta_M(\delta_{M-2})$ moves
counterclockwise (clockwise) $M-1$ times around the origin $(\delta=0)$
which is the singular point where the phase shift diverges.
This singular point in the relative rapidity space is the branch point of the
phase shift function $\theta (\lambda)$.
Thus how many times the relative rapidity winds
 around the branch point determines the corresponding increment of
 the phase shift.
When the relative rapidity $\lambda$ moves counterclockwise around
the branch point $\lambda=2i(\pi-\mu)$, the phase shift increases by
$2\pi$.
 The phase shift
$\theta(\lambda_M,\lambda_{M-1})\ (\theta(\lambda_{M-1},\lambda_{M-2}))$
increases (decreases) by $2(M-1)\pi$ due to the fluctuation of
the three string.

\midinsert
\epsfysize=10cm
$$\epsffile{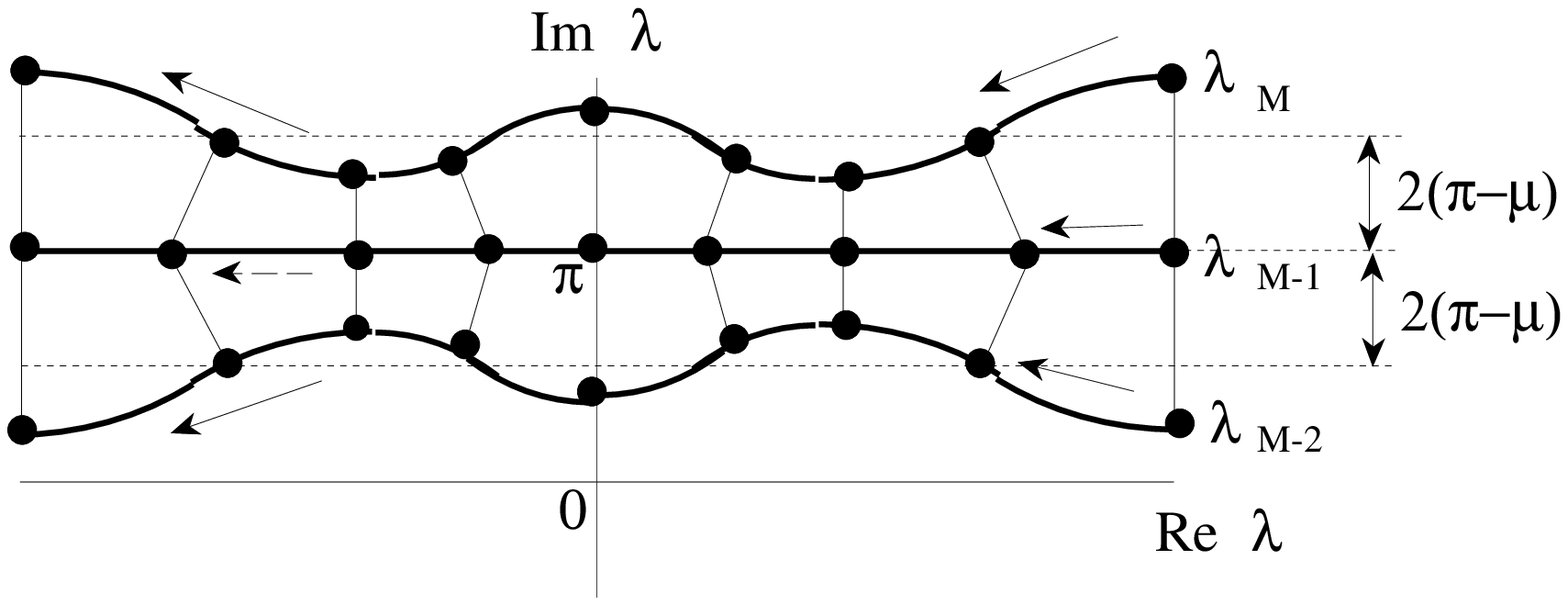}$$
\nobreak
\narrower
\singlespace
\noindent
Figure 2-4. The fluctuations of the three-string in the case of M=3 and N=6.
\medskip
\endinsert

At $\Phi=4\pi-6(\pi-\mu)$, the center of the three-string reaches
$-\infty+i\pi$.
The rapidities jump by $-{\textstyle \pi\over 3}$ and are decomposed into
$\lambda_{M-2}$ on the real axis and the two-string $(\lambda_M -
\lambda_{M-1})$ whose center is on $i\pi$ line.
At $\Phi=4\pi-4(\pi-\mu)$ the two-string is decomposed;
$\lambda_{M}$ and $\lambda_{M-1}$ jump onto the $i \pi$ line and the real axis
respectively.
Finally $\lambda_M$ jumps on to the real axis at $\Phi=4\pi-2(\pi-\mu)$.
When $\Phi=4\pi$, the set of the rapidities $\{\lambda_j^{(f)}\}$ is the same
as the set
of initial $(\Phi=0)$ rapidities $\{\lambda_j^{(i)}\}$.
The rapidities are rearranged as
$$
\eqalign{
  \lambda_M^{(f)} &= \lambda_1^{(i)},\quad \lambda_{M-1}^{(f)} =
\lambda_2^{(i)},\quad
  \lambda_{M-2}^{(f)} = \lambda_3^{(i)}, \cr
  \lambda_j^{(f)} &= \lambda_{j+3}^{(i)}\ (j=1,\ldots,M-3).}
\eqn\eq
$$

\midinsert
$$\epsffile{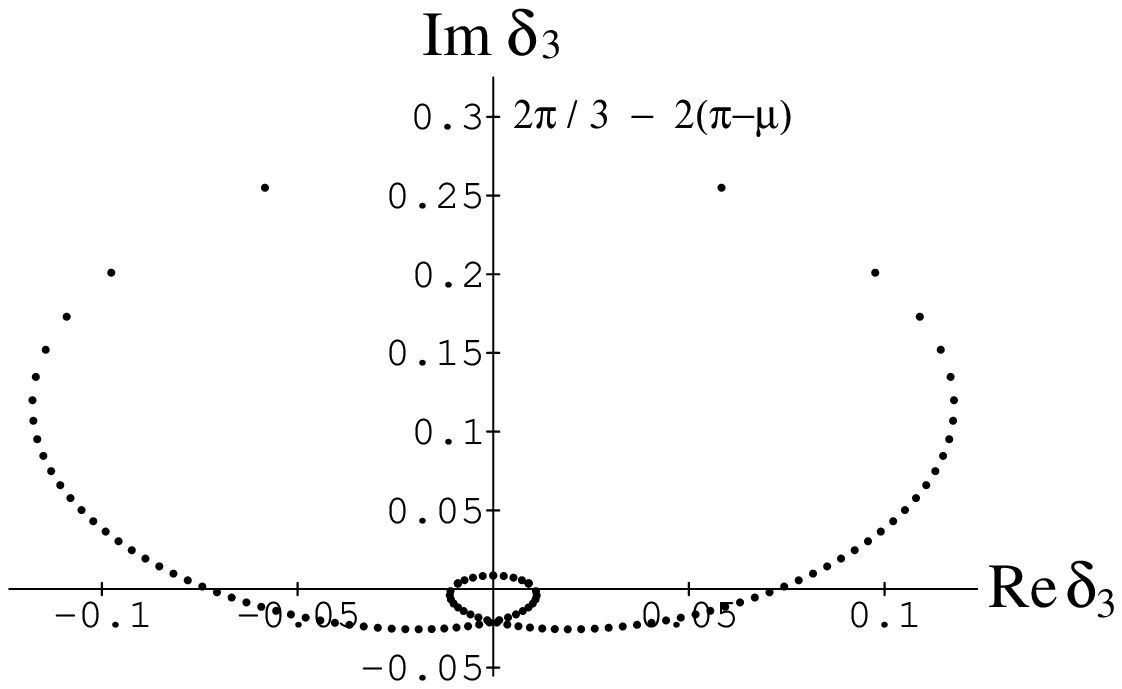}$$
\nobreak
\narrower
\singlespace
\noindent
Figure 2-5. The deviation ${\delta}_3 = (\lambda_3 - \lambda_2) - 2(\pi - \mu)$
in the case of $M=3$, $N=6$ and $\mu=5\pi /7$. When $\Phi = 6(\pi - \mu)$, the
deviation is $2\pi / 3 - 2 (\pi - \mu) \simeq 0.299$. When $\Phi$ increases
from $6(\pi - \mu )$ to $ 4\pi - 6(\pi - \mu )$, the deviation moves
counterclockwise twice around the origin.
\medskip
\endinsert

This behavior of rapidities has been checked by numerical calculation
(Fig. 2-5).
The motion of the two-string is independent of the
rapidities lying on the real axis as seen in  $\S$ 2-1.
In the case of the three-string,
 its fluctuation
is influenced by the rapidities lying on the real axis.

\midinsert
\epsfysize=10cm
$$\epsffile{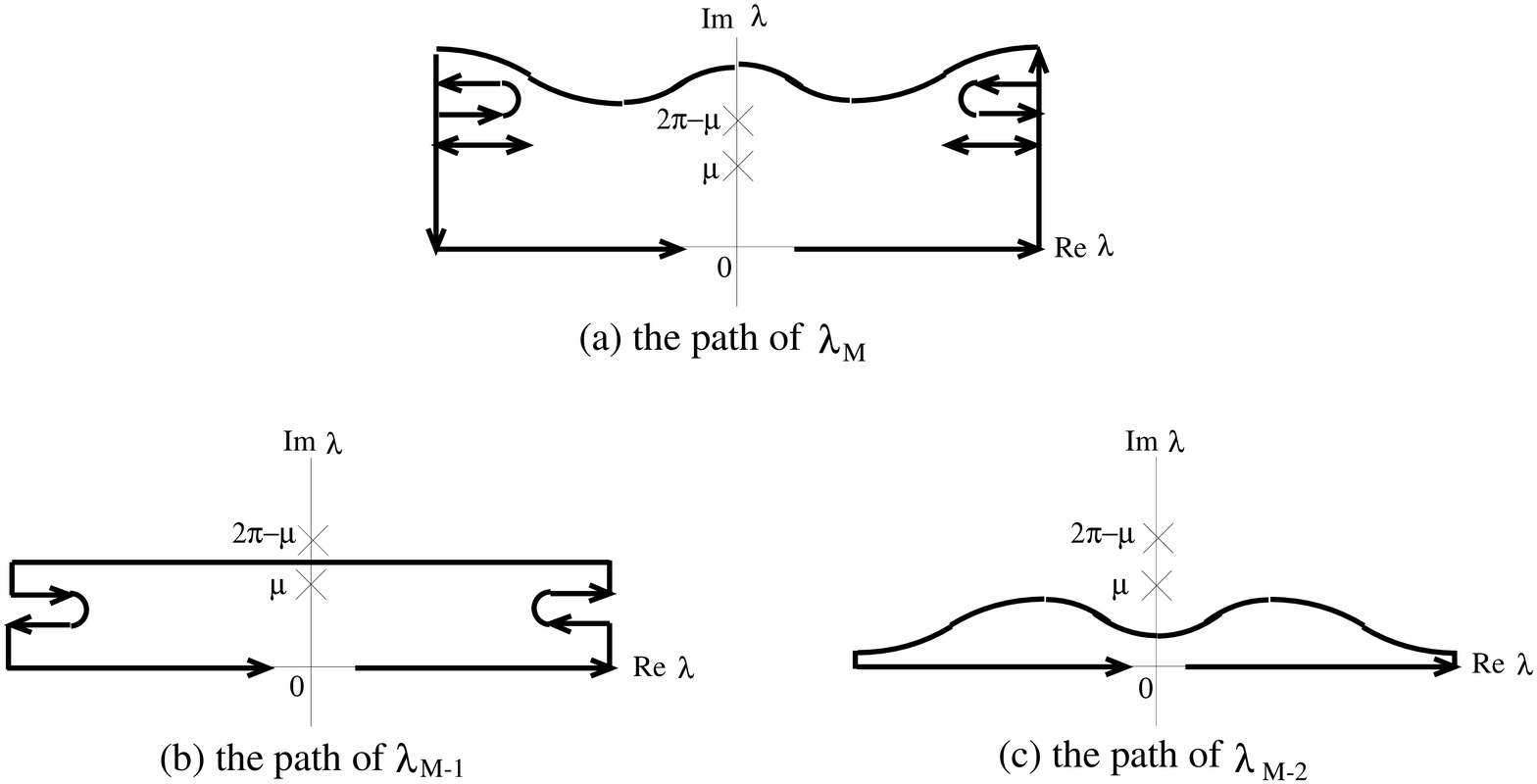}$$
\nobreak
\narrower
\singlespace
\noindent
Figure 2-6. The paths of the rapidities in the case of $2\pi / 3 < \mu < 3\pi /
4$.
\medskip
\endinsert

\midinsert
\epsfysize=9cm
$$\epsffile{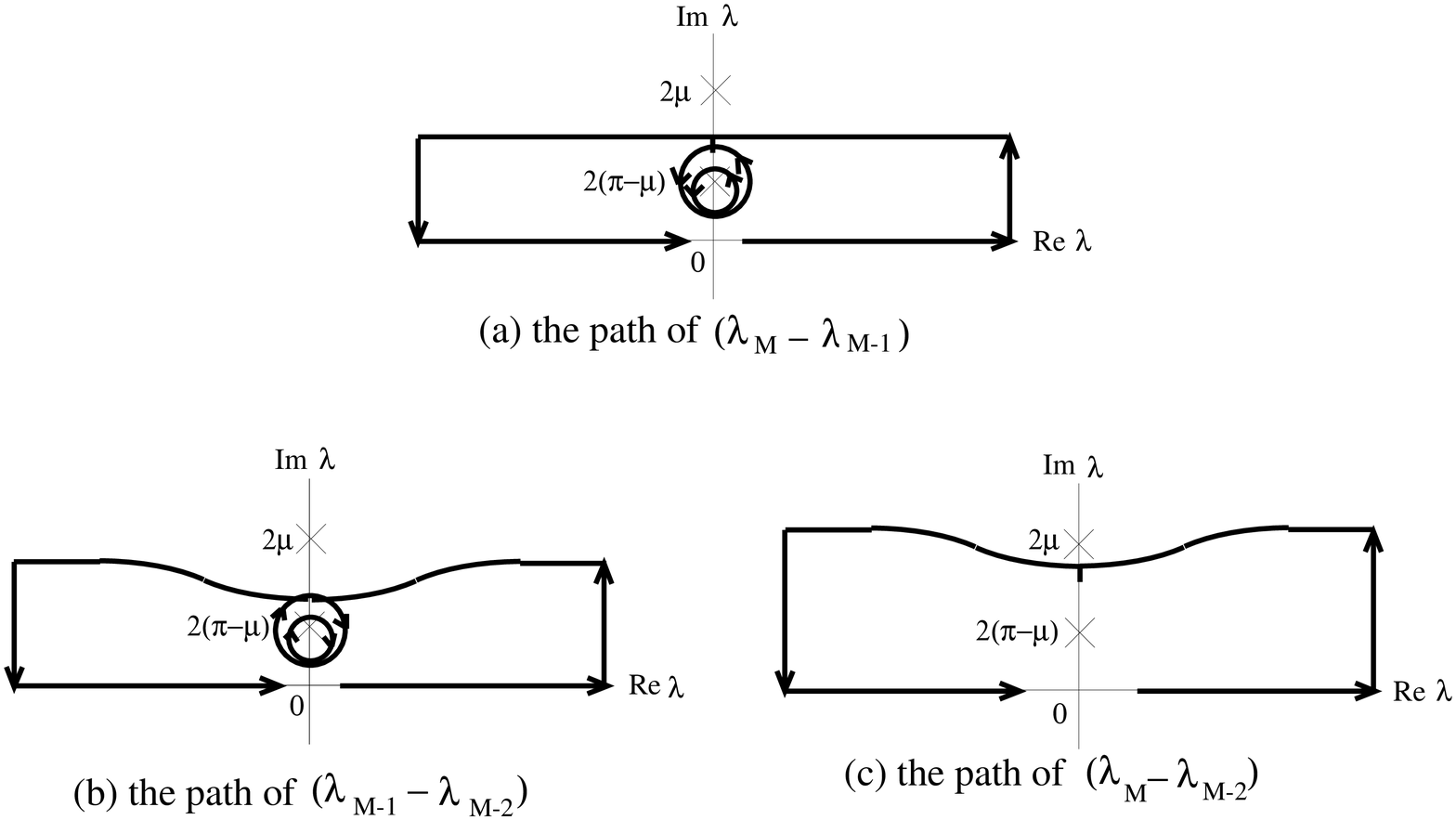}$$
\nobreak
\narrower
\singlespace
\noindent
Figure 2-7. The paths of the relative rapidities in the case of $2\pi / 3 < \mu
< 3\pi / 4$.
\medskip
\endinsert

\midinsert
\epsfysize=12cm
$$\epsffile{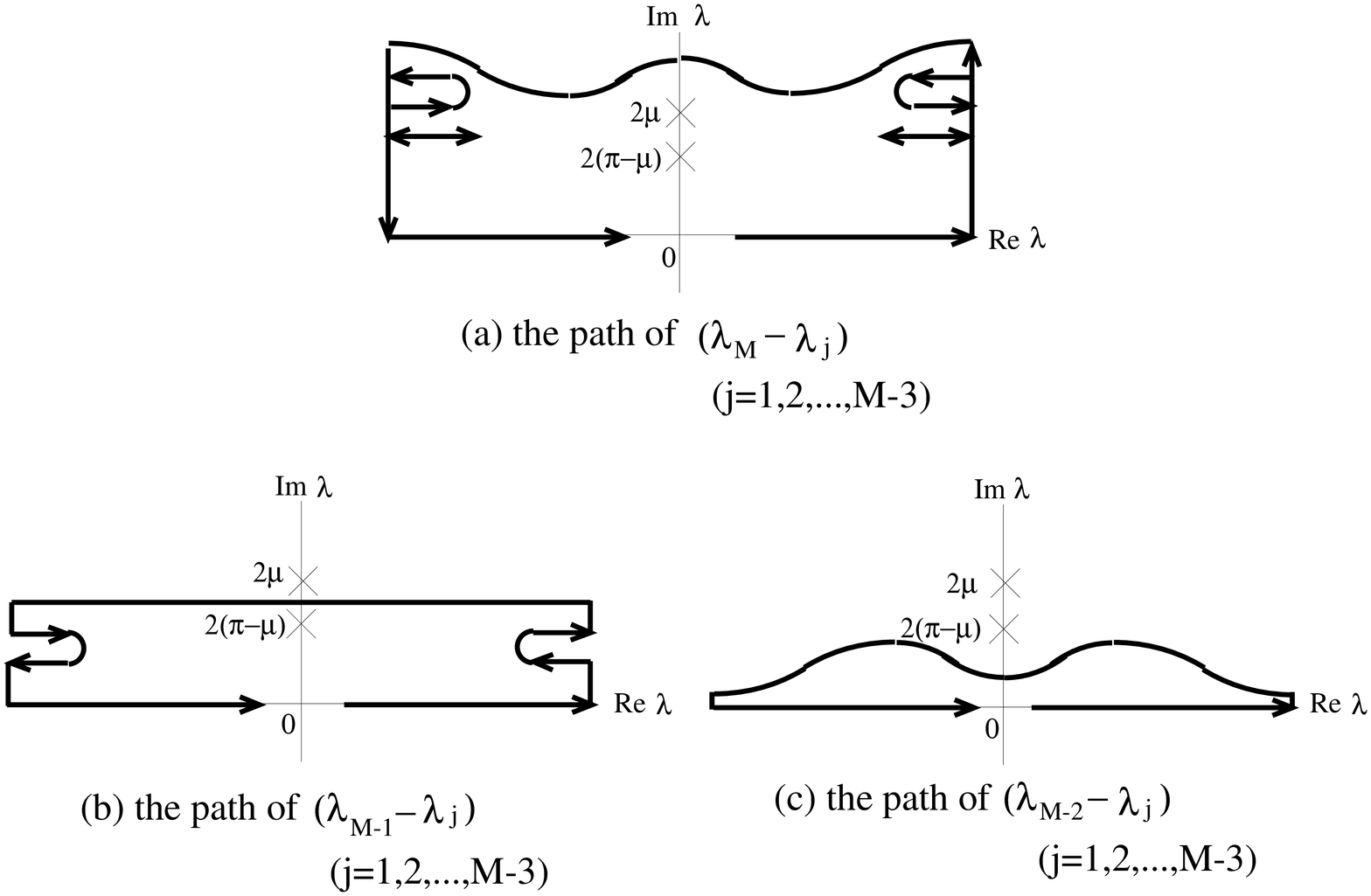}$$
\nobreak
\narrower
\singlespace
\noindent
Figure 2-8. The paths of the relative rapidities in the case of $2\pi / 3 < \mu
< 3\pi / 4$.
\medskip
\endinsert

The relations between the set $\{p_j^{(f)},\ \theta(p_j^{(f)},p_k^{(f)})\}$
and the set $\{p_j^{(i)},\ \theta(p_j^{(i)},p_k^{(i)})\}$ are determined
from the behavior of the rapidities.
The branch cut of the function $p(\lambda)$ extends from the branch point
$\lambda=i\mu$ to another branch point  $\lambda=i(2\pi-\mu)$.
Only the path of $\lambda_{M-1}$ crosses the cut (Fig. 2-6),  so that the
relations of momenta are
$$
\eqalign{
  p_M^{(f)} &= p_1^{(i)},\quad p_{M-1}^{(f)} = p_2^{(i)} + 2\pi,\quad
  p_{M-2}^{(f)} = p_3^{(i)}, \cr
  p_j^{(f)} &= p_{j+3}^{(i)}\ (j=1,\ldots,M-3).}
\eqn\IIIpa
$$
The cut of $\theta(\lambda)$ extends from the branch point
$\lambda=2i(\pi-\mu)$ to another branch point $\lambda=2i\mu$.
The path of the relative rapidity $\lambda_M-\lambda_{M-1}$,
 and those of
$\lambda_{M-1}-\lambda_{M-2}$ and $\lambda_M-\lambda_{M-2}$
cross the cut $1+(M-1)$ times, $1-(M-1)$ times and once respectively
(Fig. 2-7).
 We have, therefore,
$$
\eqalign{
  \theta(\lambda_M^{(f)},    \lambda_{M-1}^{(f)}) &=
  \theta(\lambda_1^{(i)},\lambda_2^{(i)}) + 2M\pi,\cr
  \theta(\lambda_{M-1}^{(f)},\lambda_{M-2}^{(f)}) &=
  \theta(\lambda_2^{(i)},\lambda_3^{(i)}) - 2(M-2)\pi,\cr
  \theta(\lambda_M^{(f)},    \lambda_{M-2}^{(f)}) &=
  \theta(\lambda_1^{(i)},\lambda_3^{(i)}) + 2\pi.}
\eqn\eq
$$
For the other phase shifts, only the path
$\lambda_{M-1}-\lambda_k\ (k=1,\ldots,M-3)$ crosses the cut once (Fig. 2-8).
 We have
$$
\eqalign{
  \theta(\lambda_M^{(f)},    \lambda_k^{(f)}) &=
  \theta(\lambda_1^{(i)},\lambda_{k+3}^{(i)}),\cr
  \theta(\lambda_{M-1}^{(f)},\lambda_k^{(f)}) &=
  \theta(\lambda_2^{(i)},\lambda_{k+3}^{(i)}) + 2\pi,\cr
  \theta(\lambda_{M-2}^{(f)},\lambda_k^{(f)}) &=
  \theta(\lambda_3^{(i)},\lambda_{k+3}^{(i)}),\cr
  \theta(\lambda_j^{(f)},    \lambda_k^{(f)}) &=
  \theta(\lambda_{j+3}^{(i)},\lambda_{k+3}^{(i)}), \cr
  & \ (j,k=1,\ldots,M-3).}
\eqn\IIIpb
$$

Finally, let us check  the consistency between the relations \IIIpa --\IIIpb\
and the twisted boundary conditions \trans.
The boundary conditions  for the final state are
$$
\eqalign{
  2Mp_j^{(f)}    +\sum_{k\neq j}  \theta(p_j^{(f)},    p_k^{(f)})
&=  2\pi\left( {2j-M-1\over 2} \right) + 4\pi,\cr
&\ (j=1,\ldots,M ).}
\eqn\eq
$$
These equations are rewritten in terms of the initial set by using
\IIIpa --\IIIpb,
$$
   2Mp_1^{(i)}+
  \left(
    \sum_{k\neq 1}\theta(p_1^{(i)},p_{j+3}^{(i)})+2M\pi+2\pi
  \right)
  = 2\pi \left( {M-1\over 2}\right) + 4\pi,
\eqn\eq
$$
$$
   2M(p_2^{(i)}+2\pi)+
  \left(
    \sum_{k\neq 2}\theta(p_2^{(i)},p_{j+3}^{(i)})-2M\pi-2(M-2)\pi+2(M-3)\pi
  \right)
  = 2\pi \left( {M-3\over 2}\right) + 4\pi,
\eqn\eq
$$
$$
   2Mp_3^{(i)}+
  \left(
    \sum_{k\neq 3}\theta(p_3^{(i)},p_{j+3}^{(i)})+2(M-2)\pi-2\pi
  \right)
  = 2\pi \left( {M-5\over 2}\right) + 4\pi,
\eqn\eq
$$

$$
   2Mp_{j+3}^{(i)}+
  \left(
    \sum_{k\neq j+3}\theta(p_{j+3}^{(i)},p_{j+3}^{(i)})-2\pi
  \right)
  = 2\pi \left( {2j-M-1\over 2}\right) + 4\pi,\ \
  \ (j=1,\ldots,M-3).
\eqn\eq
$$
They are equivalent to the boundary conditions for the initial state:
$$
2Mp_1+\sum_{k\neq 1}\theta(p_1^{(i)},p_k^{(i)}) =
2\pi\left(-{M-1\over 2}\right) + 0,
\eqn\eq
$$
$$
2Mp_2+\sum_{k\neq 2}\theta(p_2^{(i)},p_k^{(i)}) =
2\pi\left(-{M-3\over 2}\right) + 0,
\eqn\eq
$$
$$
2Mp_3+\sum_{k\neq 3}\theta(p_3^{(i)},p_k^{(i)}) =
2\pi\left(-{M-5\over 2}\right) + 0,
\eqn\eq
$$
$$
\eqalign{
  2Mp_{j+3}+\sum_{k\neq j+3}\theta(p_{j+3}^{(i)},p_k^{(i)}) &=
  2\pi\left(-{2{j+3}-M-1\over 2}\right) + 0, \cr
  & \ (j=1,\ldots,M-3).}
\eqn\eq
$$

\section{ Generalization to the case of $n$-string}

  We have found the collision of the rapidities
   forming the two-string into the singularities
  as well as the fluctuation of the three-string around its center
 in the previous subsections.
 It is possible to generalize  these behaviors to the case of  $n$-string.
   Let us describe the motion of the rapidities we have found in the region
${\rm cos}({\pi \over n})< \Delta < {\rm cos}({\pi \over n+1})$
$({n-1 \over n}\pi<\mu<{n \over n+1}\pi)$.
  At $\Phi=0$,  all rapidities are symmetrically arranged on the real axis
   with respect to the origin.
  At $\Phi=2j(\pi-\mu)$ for  $(j=1,2,\cdots,n)$,  a group of $j$ rapidities at
$Re(\lambda)=\infty$  jumps: a $j$-string is formed from
 a $(j-1)$-string and one string .
  The outcome of this series of  events is the formation of the $n$-string
 at $ \Phi=2n(\pi-\mu)$.  After this, the center of
 the $n$-string moves backwards on the $i\pi$ line from
$\infty+i\pi$ to $-\infty+i\pi$  as $\Phi$ increases from
  $2n(\pi-\mu)$ to $4\pi-2(\pi-\mu)$.   The $n$-string itself fluctuates
  during this period
  as we have seen in the case of the three-string. (We will describe this
   shortly for  the case $n$ even and   for the case $n$ odd separately.)
  As $\Phi$ increases further,  a separation of a j-string into a (j-1)-string
  and one string on the real axis occurs at $ \Phi=4\pi-2j(\pi-\mu)$
 for $j=n,n-1,\cdots,1$.
  Finally at $\Phi=4\pi$, all rapidities are again symmetrically arranged
  on the real axis with respect to the origin.

   Let us now describe the fluctuation of the $n$-string.   In the case of
  $n$ even,  the two rapidities in the middle, namely,
$\lambda_{M-{n\over 2}+1}$ and $\lambda_{M-{n \over 2}}$ of the $n$-string
   collide into the two branch points  of $p(\lambda)$
 at $\Phi=2\pi$:
$$
\lambda_{M-{n \over 2}+1}=2\pi-\mu, \ \ \ \ \ \
\lambda_{M-{n\over 2}}=\mu.
\eqn\nebrpt
$$
 The other rapidities fluctuate: the deviation
$$
\delta_{M-j+1}\equiv(\lambda_{M-j+1}-\lambda_{M-j})-2(\pi-\mu)\, \ \ \
(j=1,2,\cdots,{n \over 2}-1)
\eqn\nedeva
$$
   moves counterclockwise  $(M-n+2j)$  times  around
$\delta=0$,  which is the branch point of $\theta(\lambda)$.
 The deviation
$$
\delta_{M-n+j}\equiv(\lambda_{M-n+j}-\lambda_{M-n+j+1})-2(\pi-\mu)
\, \ \ \ (j=1,2,\cdots,{n \over 2}-1)
\eqn\nedevb
$$
 moves clockwise  $(M-n+2j)$  times   around
$\delta=0$.

  In the case of $n$ odd,   the rapidity in the middle
$\lambda_{M-{n-1 \over 2}}$  moves on the
$i\pi$ line  while the other rapidities  fluctuate:  the deviation
$$
\delta_{M-j+1}\equiv(\lambda_{M-j+1}-\lambda_{M-j})-2(\pi-\mu) \, \ \ \
(j=1,2,\cdots,{n-1 \over 2})
\eqn\nodeva
$$
   moves counterclockwise  $(M-n+2j)$  times  around
$\delta=0$  and  the deviation
$$
\delta_{M-n+j}\equiv(\lambda_{M-n+j}-\lambda_{M-n+j+1})-2(\pi-\mu) \, \ \ \
(j=1,2,\cdots,{n-1 \over 2})
\eqn\nodevb
$$
 moves clockwise  $(M-n+2j)$  times   around
$\delta=0$.
  Only the rapidity  $\lambda_{M-{n-1 \over 2}}$ crosses the cut.

The rationale for this behavior of the rapidities lies
 in  the fact that they pass a set of consistency conditions,
 which is found to be very stringent.
 Let us denote   by $\{\lambda^{(i)}_j\}$ a set of  initial rapidities
  at $\Phi=-2\pi$ and by $\{\lambda^{(f)}_j\}$ a set of final rapidities
 at $\Phi=2\pi$.  We have
$$
\eqalign{\lambda^{(f)}_{M-j+1} &=\lambda^{(i)}_j \ \ (j=1,2,\cdots,n), \cr
\lambda^{(f)}_{j}&=\lambda^{(i)}_{j+n} \ \ (j=1,2,\cdots,M-n) \cr}.
\eqn\mrpfi
$$
  From the paths
 of the rapidities described above, we obtain a set of relations
  for the momenta
$$
p(\lambda^{(f)}_{M-j+1})=p(\lambda^{(i)}_j), \ \ \
p(\lambda^{(f)}_{M-n+j})=p(\lambda^{(i)}_{n-j+1})\ \ \
(j=1,2,\cdots,\left[{n-1 \over 2}\right]).
$$
$$
p(\lambda^{(f)}_{M-{n\over 2}+1})=p(\lambda^{(i)}_{n \over 2})+\pi, \ \ \
p(\lambda^{(f)}_{M-{n\over 2}})
=p(\lambda^{(i)}_{{n\over 2}+1})+\pi \ \ \ (n:{\rm even}).$$
$$
p(\lambda^{(f)}_{M-{n-1\over 2}})=P(\lambda^{(i)}_{n+1 \over 2})+2\pi
\ \ \ (n:{\rm odd}).
$$
$$
p(\lambda^{(f)}_j)= p(\lambda^{(i)}_{j+n}) \ \ \ (j=1,2,\cdots,M-n).
\eqn\relmom
$$
 From the paths of the relative rapidities, we obtain a set of relations for
  the phase shifts:
$$
\theta(\lambda^{(f)}_{M-j+1},\lambda^{(f)}_{M-j})=
\theta(\lambda^{(i)}_j,\lambda^{(i)}_{j+1})+2\pi\{1+(M-n+2j)\},
$$
$$
(j=1,2,\cdots,\left[{n-1 \over 2}\right]).
\eqn\npsa
$$
$$
\theta(\lambda^{(f)}_{M-n+j},\lambda^{(f)}_{M-n+j+1})=
\theta(\lambda^{(i)}_{n-j+1},\lambda^{(i)}_j)+2\pi\{1+(M-n+2j)\},
$$
$$
(j=1,2,\cdots,\left[{n-1 \over 2}\right]).
\eqn\npsb
$$
$$
\theta(\lambda^{(f)}_{M-{n\over 2}+1},\lambda^{(f)}_{M-{n\over 2}})=
\theta(\lambda^{(i)}_{{n\over 2}-1},\lambda^{(i)}_{n\over 2})+2\pi,
$$
$$
\theta(\lambda^{(f)}_{M-j+1},\lambda^{(f)}_{M-k+1})=
\theta(\lambda^{(i)}_j,\lambda^{(i)}_k)+2\pi \ \ \ \
(1\leq j<k\leq n \ {\rm and} \ k-j\geq 2),
$$
$$
\theta(\lambda^{(f)}_M,\lambda^{(f)}_k)=
\theta(\lambda^{(i)}_1,\lambda^{(i)}_{k+n}) \ \ \ \ (k=1,\cdots,M-n),
$$
$$
\theta(\lambda^{(f)}_{M-n+1},\lambda^{(f)}_k)=
\theta(\lambda^{(i)}_n,\lambda^{(i)}_{k+n}) \ \ \ \ (k=1,\cdots,M-n),
$$
$$
\theta(\lambda^{(f)}_{M-j+1},\lambda^{(f)}_k)=
\theta(\lambda^{(i)}_j,\lambda^{(i)}_{k+n})+2\pi \ \ \ \
(2\leq j\leq n-1,k=1,\cdots,M-n),
$$
$$
\theta(\lambda^{(f)}_j,\lambda^{(f)}_k)=
\theta(\lambda^{(i)}_{j+n},\lambda^{(i)}_{k+n}) \ \ \ \
(j=1,\cdots,M-n,k=1,\cdots,M-n).
\eqn\npsc
$$
 In the first and the second formulas,  the factor $\pm(M-n+2j)$ represents
 a contribution due to the fluctuation of the $n$-string.  The third formula is
 for the case of $n$ even only.
  Following the same procedure as described in \S 2.1 and\S 2.2,   and
 using the relations given above, we have
  checked  that the twisted boundary conditions for the set of the initial
 rapidities $\{\lambda^{(i)}_j\}$  follow from
  those for the set of the final rapidities
$\{\lambda^{(f)}_j\}$. The proof is too long and tedious to be presented here.

  This consistency check is in fact very stringent. We are convinced
 that the picture of  the $n$-string fluctuations we have found in this
  paper has now been  established  by this.

\endpage


\chapter{ Results in the Regime $\Delta \leq -1$ }

In this section, we study   the ground state properties
 mentioned in the introduction of the XXZ model
 in  the regime $\Delta \leq -1$. Throughout this regime, the single particle
  energy is kept non-positive.

\section{ the regime  $\Delta<-1$}

In this regime, the momentum is parametrized in terms of
the rapidity as
$$
p(\lambda) = -i{\rm ln}
\left[
  -{{{\rm sinh}{1\over 2}(i\lambda+\mu)}
    \over{{\rm sinh}{1\over 2}(i\lambda-\mu)}}
\right],
\eqn\eq
$$
where $\Delta=-{\rm cosh}\mu$. As the momentum increases from zero
to $\pi$ (from $\pi$ to $2\pi$), the rapidity increases from zero
to $\pi$ (from $\pi$ to $2\pi$).
The particle-particle phase shift becomes
$$
\theta(\lambda_1,\lambda_2) = i{\rm ln}
\left[
  -{{{\rm sinh}{1\over 2}(i(\lambda_1-\lambda_2)+2\mu)}
    \over{{\rm sinh}{1\over 2}(i(\lambda_1-\lambda_2)-2\mu)}}
\right].
\eqn\eq
$$
As the relative rapidity increases from zero to $\pi$ (from $\pi$ to
$2\pi$), the phase shift decreases from zero to $-\pi$ (from $-\pi$
to $-2\pi$).
We plot the momentum $p(\lambda)$ as a function of the
rapidity and the phase shift $\theta (\lambda)$ as a function of the relative
rapidity $\lambda $ in Fig. 3-1.
The total energy is expressed as
$$
E = {N\over 2} {\rm cosh}\mu + \sum_{j=1}^M
    {{-2{\rm sinh}^2\mu}\over{{\rm cosh}\mu-{\rm cos}\lambda_j}}.
\eqn\eq
$$
The single-particle energy
$$
 \varepsilon_j = {{-2{\rm sinh}^2\mu}
                \over{{\rm cosh}\mu-{\rm cos}\lambda_j}}
\eqn\asenergy
$$
is negative for the real $\lambda_j$.

\midinsert
$$\epsffile{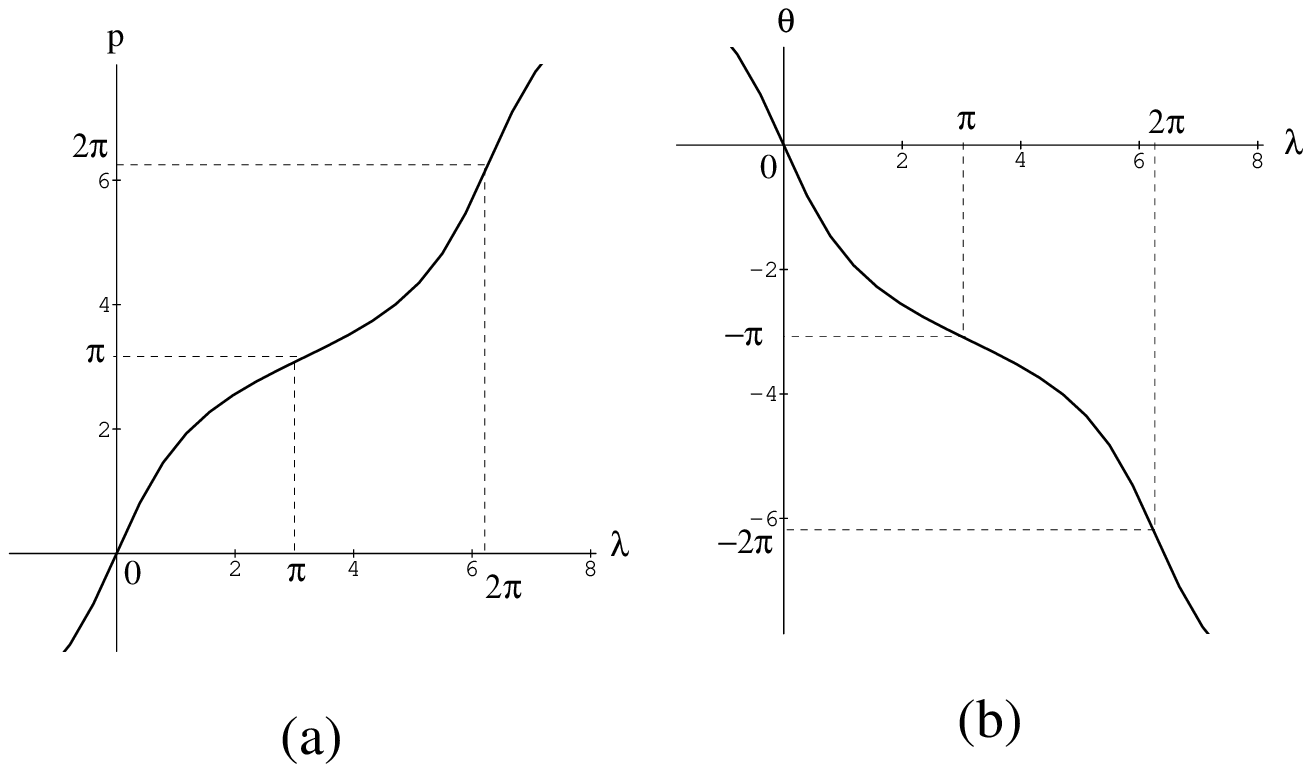}$$
\nobreak
\narrower
\singlespace
\noindent
Figure 3-1.  The function for the momentum and the function for the phase-shift
function in the case of $\Delta <-1$.
\medskip
\endinsert

Let us consider the  process associated with the change
of the twist angle $\Phi$ from zero to $4\pi$.
The motion of the rapidities is determined from equation
\trans\ with \Betheqn.
When $\Phi=0$, all the rapidities $\lambda_j$
are real and symmetric with respect to the origin of the rapidity
space ($\{\lambda_j\} = \{-\lambda_j\}$) within the Brillouin zone
$-\pi<\lambda<\pi$.
At $\Phi=2\pi$, the largest rapidity $\lambda_M$  (or one-string)
 reaches $\pi$ and
the rest of them are symmetric with respect to the origin
 of the rapidity plane $\lambda = 0$.
This is an excited state for the periodic boundary condition.
At $\Phi=4\pi$ we will denote the configuration of $\lambda_j$ by
$\{\lambda_j^{(f)}\}$. It is related to the original ground state
configuration by
$$
\lambda_M^{(f)}=\lambda_1^{(i)}+2\pi,\qquad
\lambda_j^{(f)}=\lambda_{j+1}^{(i)} \quad (j\neq M).
\eqn\appa
$$
This  leads to the relations
$$
p_M^{(f)} = p_1^{(i)} + 2\pi,\qquad p_j^{(f)} = p_{j+1}^{(i)}\quad (j\neq M),
\eqn\appb
$$
$$
\theta(p_M^{(f)},p_j^{(f)})=\theta(p_1^{(i)},p_{j+1}^{(i)})-2\pi,
\eqn\appc
$$
$$
\theta(p_j^{(f)},p_k^{(f)})=\theta(p_{j+1}^{(i)},p_{k+1}^{(i)})\quad
(j\neq M,k\neq M),
$$
where $p_j^{(f)}=p(\lambda_j^{(f)})$ and $p_j^{(i)}=p(\lambda_j^{(i)})$.
The set of the momenta for the final state coincides with
  the one for the initial state up to $2\pi$.
Accordingly the system returns to the ground state at $\Phi=4\pi$.

We check briefly the consistency between the above equation and
the twisted boundary condition \trans\ at $\Phi=0,2\pi,4\pi$.
The twisted boundary condition for $\Phi=0$ is
$$
2Mp(\lambda_j^{(i)})+\sum_{k\neq j}\theta(\lambda_j^{(i)},\lambda_k^{(i)})
= 2\pi{2j-M-1\over 2}.
\eqn\eq
$$
As the $p(\lambda)$ and $\theta(\lambda)$ are odd functions,  the
above equation becomes
$$
2Mp(-\lambda_j^{(i)})+\sum_{k\neq j}\theta(-\lambda_j^{(i)},-\lambda_k^{(i)})
= 2\pi{-M+2(M-j+1)-1\over 2}.
\eqn\eq
$$
Comparing this with the twisted boundary condition for $p(\lambda_{M-j+1}^i)$
$$
2Mp(\lambda_{M-j+1}^{(i)})+\sum_{k\neq j}
\theta(\lambda_{M-j+1}^i,\lambda_{M-k+1}^i)
= 2\pi{-M+2(M-j+1)-1\over 2},
\eqn\eq
$$
we find that the configuration of the rapidities is symmetric
$\lambda_{M-j+1}^{(i)}=-\lambda_j^{(i)}$.
In the case of $\Phi=2\pi$, we can derive $\lambda_M=\pi$ and
$\lambda_{M-j}=-\lambda_j\ (j\neq M)$ in the same way.
When $\Phi=4\pi$, the twisted boundary condition for $p(\lambda_j^{(f)})$
is
$$
2Mp(\lambda_j^{(f)})+\sum_{k\neq j}\theta(\lambda_j^{(f)},\lambda_k^{(f)})
= 2\pi{2j-M-1\over 2} + 4\pi.
\eqn\eq
$$
Using Eqs. \appb\ and  \appc, the left hand side of the equation
is rewritten in terms of $\lambda_j^{(i)}$,
$$
2Mp(\lambda_{j+1}^{(i)})+ \left( \sum_{k\neq {j+1}}
  \theta(\lambda_{j+1}^{(i)},\lambda_k^{(i)})+2\pi
\right)
= 2\pi{2j-M-1\over 2} + 4\pi,
\eqn\atransa
$$
for $j\neq M$ and
$$
2Mp(\lambda_1^{(i)})+\sum_{k\neq 1}
\left(
  \theta(\lambda_1^{(i)},\lambda_k^{(i)})-2\pi
\right)
= 2\pi{M-1\over 2} + 4\pi,
\eqn\atransb
$$
where we used the relation
$\theta(\lambda_j,\lambda_k)=-\theta(\lambda_k,\lambda_j)$.
Eqs. \atransa\ and \atransb\ are equivalent to the boundary
conditions for the initial state
$$
2Mp(\lambda_{j+1}^{(i)})+\sum_{k\neq {j+1}}
\theta(\lambda_{j+1}^{(i)},\lambda_k^{(i)})
= 2\pi{2(j+1)-M-1\over 2} + 0,
\eqn\eq
$$
$$
2Mp(\lambda_1^{(i)})+\sum_{k\neq 1}\theta(\lambda_1^{(i)},\lambda_k^{(i)})
= 2\pi
\left(
  -{M-1\over 2}
\right)
+ 0.
\eqn\eq
$$
We confirm the motion of the rapidities for the case of $M=2$ and $N=4$
 in Appendix A.
Observe that the single-particle energy \asenergy\ remains
negative during the  process.

\section{  the  $\Delta= -1$ case}

Let us now study the ground state in the critical
coupling case $\Delta= -1$.
When $\Delta= -1$, the momentum $p$ is parametrized in terms of
rapidity $\lambda$ by
$$
p(\lambda) = 2\,{\rm arctan}(2\lambda).
\eqn\eq
$$
When the momentum increases from zero to $\pi$, the rapidity goes from
zero to $+\infty$.
As the momentum increases still more the rapidity jumps from $+\infty$ to
$-\infty$.
When the momentum increase from $\pi$ to $2\pi$, the rapidity goes from
$-\infty$ to zero.
The phase shift and total energy become
$$
\theta(\lambda_1,\lambda_2) = -2\,{\rm arctan}(\lambda_1-\lambda_2)
\eqn\eq
$$
and
$$
E={N\over 2}+\sum_{j=1}^M{-1\over 1+4\lambda_j^2},
\eqn\eq
$$
respectively.
The single-particle energy
$$
\varepsilon_j = {-1\over 1+4\lambda_j^2}
\eqn\ccsenergy
$$
is negative for real $\lambda_j$ and zero for $\lambda_j=\pm\infty$.

Let us describe the process associated with the change of
twist angle $\Phi$.
When $\Phi=0$, all the rapidities $\lambda_j$ are real and symmetric
with respect to the origin of the rapidity space.
If we set $\Phi=2\pi$, the largest rapidity $\lambda_M$ goes to plus infinity
and jumps to minus infinity.
At $\Phi=4\pi$ the configuration of rapidities
$\{\lambda_j\}$ comes back to the initial configuration as follows
$$
\lambda_M^{(f)} = \lambda_1^{(i)},\qquad
\lambda_j^{(f)} = \lambda_{j+1}^{(i)}\quad (j\neq M).
\eqn\ccpa
$$
This leads to the relations
$$
p_M^{(f)} = p_1^{(i)} + 2\pi,\qquad p_j^{(f)} = p_{j+1}^{(i)}\quad (j\neq M),
\eqn\ccpb
$$
$$
\eqalign{
\theta(p_M^{(f)},p_j^{(f)}) &= \theta(p_1^{(i)},p_{j+1}^{(i)}) -2\pi, \cr
\theta(p_j^{(f)},p_k^{(f)}) &= \theta(p_{j+1}^{(i)},p_{k+1}^{(i)}). }
\eqn\ccpc
$$
This is the same as the relations \appb\ and \appc\ in the $\Delta < -1$ case.
The  period of the
system  is again $\Delta\Phi = 4\pi$.

We have found that the motion of the rapidities, the period, the final
  configuration of the momenta and that of the phase shift functions
  in the case $ \Delta \leq  -1$
  are the same as  those found in the case $ -1 < \Delta < 0$.
  We have, therefore, established the continuity of the two regimes.
  This can be heuristically explained  by the structure of the
 degeneracies of the levels, which we illustrate  for the  case
   $M=2,N=4$ in Fig. 3-2.  The ground state at $\Phi=0$
  never gets degenerated with the zero energy levels  as long
   as $\Delta < 0$.  This is, however, seen for $\Delta \geq 0$.

\midinsert
$$\epsffile{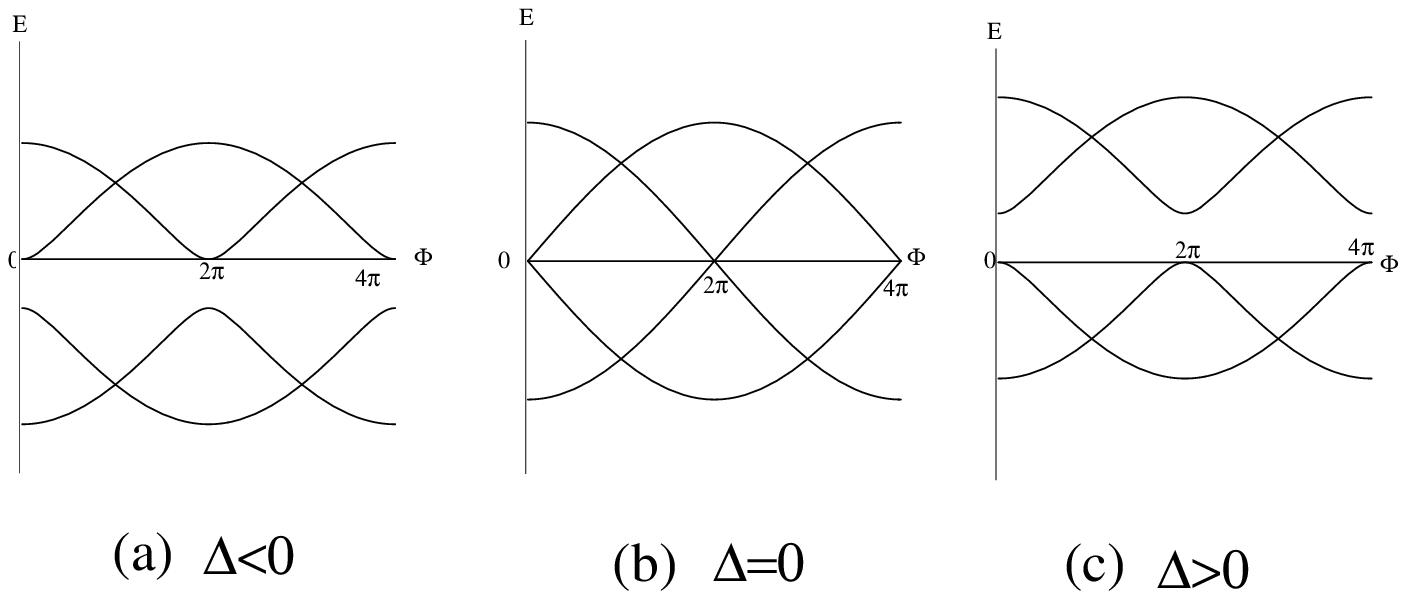}$$
\nobreak
\narrower
\singlespace
\noindent
Figure 3-2. The energy levels of XXZ model in M=2 and N=4 case. The
eigenenergies are $E = \Delta \pm \sqrt{\Delta^2 + 8 \cos^2 \Phi/4},
 \Delta \pm \sqrt{\Delta^2 + 8 \sin^2 \Phi/4},0,0 $ which we derive in
 Appendix A.
\medskip
\endinsert

\endpage

\chapter{ The Berry Phase }

  Let us now turn to the discussion of the Berry phase
   associated with the twist angle $\Phi(t)$, which is now considered to
   vary with time.
   We  are interested in  computing  the
 nontrivial geometrical phase $\gamma$ appearing in the
  wave function
$$
|\psi(\Phi(t),t)\bigr> = {\rm exp}
\left\{
  -i\int_0^t\!\!E_0(\Phi(t))dt
\right\}
{\rm exp}(i\gamma(t)) |\psi_0(\Phi(t))\bigr>.
\eqn\solSch
$$
 The substitution of this into the Schrodinger equation provides a formula
$$
\gamma(t) = {\rm Re}
\left[
  i\int_0^t\!\!dt
  {\bigl<\psi_0(\Phi(t))|{\partial\over\partial \Phi}|\psi_0(\Phi(t))\bigr>
   \over
   \bigl<\psi_0(\Phi(t))|\psi_0(\Phi(t))\bigr>} {d \Phi \over d t}
\right].
\eqn\eq
$$
In our consideration, the initial and the final twist angles are taken
 to be $\Phi(0)=-2\pi$ and $\Phi(t) = +2\pi$ respectively.
After a change of variables, we find
$$
\gamma = {\rm Re}
\left[
  i\int_{-2\pi}^{2\pi}\!\!d\Phi
  {\bigl<\psi_0(\Phi)|{\partial\over\partial\Phi}|\psi_0(\Phi)\bigr>
   \over
   \bigl<\psi_0(\Phi)|\psi_0(\Phi)\bigr>}
\right].
\eqn\Berryph
$$
  We compute this quantity in different regions.

\section{ $\Delta  \leq -1$ }

We take the half-integer coordinate
$x_j=-{N-1\over 2},-{N-3\over 2},\ldots,{N-1\over 2},$
where $N=2M$.
We first examine the relation between the wave function
$\chi(\Phi)$ of the initial state at $\Phi=-2\pi$ and that of
 the final state at $\Phi=2\pi$.
  We  first want to establish   ( see \Bethest)
$$
|\chi(-2\pi)\bigr> = -|\chi(2\pi)\bigr>.
\eqn\esif
$$

By using the permutation
$$
Q'_a =
\left\{
  \eqalign{ Q_a+1\quad & ({\rm for}\  Q_a\neq M), \cr
            1    \quad & ({\rm for}\  Q_a  =  M),}
\right.
\eqn\eq
$$
   and the relations \appa, \appb\ and \appc\
  we  find the difference in the phase factors  from the initial
state  to the final state :
$$
(-1)^{|Q|} = (-1)^{|Q'|}\times (-1)^{M-1}, \eqn\eq
$$
$$
{\rm exp} \left( i\sum_{a=1}^Mx_ap_{Q_a}^{(f)}  \right) =
{\rm exp} \left( i\sum_{a=1}^Mx_ap_{Q'_a}^{(i)} \right)
\times (-1),
\eqn\eq
$$
$$
\eqalign{  {\rm exp}
  \left(
    {i\over 2}\sum_{a<b}\theta(p_{Q_b}^{(i)},p_{Q_a}^{(i)})
    \epsilon(x_b-x_a)
  \right) \
 = {\rm exp}
  \left(
    {i\over 2}\sum_{a<b}\theta(p_{Q'_b}^{(f)},p_{Q'_a}^{(f)})
    \epsilon(x_b-x_a)
  \right)
  \times (-1)^{M-1} } .
\eqn\eq
$$
Thus under the full period   $\Phi = 4\pi$, $\chi$ gets multiplied by
  a factor
   $ (-1)^{M-1}(-1)(-1)^{M-1} = -1$, which is what we wanted to show.

Identifying
$$
|\psi_0(\Phi)\bigr> = {\rm exp}(-i\Phi/4)|\chi(\Phi)\bigr>,
\eqn\eq
$$
  We find
$$
\gamma = \pi + {\rm Re}
\left[
  i\int_{-2\pi}^{2\pi}\!\!d\Phi\sum_{\{x_i\}}
  {\chi(x_1,\ldots,x_M)^*{\partial\over\partial\Phi}
   \chi(x_1,\ldots,x_M) \over
   \chi(x_1,\ldots,x_M)^*\chi(x_1,\ldots,x_M)}
\right].
\eqn\Berryform
$$
The summation $\sum_{\{x_i\}}$ represents $\prod_{i=1}^M\sum_{x_i}$,
where each $x_i$ runs over the region of half integer values.
The function of $\Phi$
$$
\sum_{\{x_i\}}
{\chi(x_1,\ldots,x_M|p_1(\Phi),\ldots p_M(\Phi))^*
{\partial\over\partial\Phi}
\chi(x_1,\ldots,x_M|p_1(\Phi),\ldots p_M(\Phi)) \over
\|\chi(\Phi)\|^2}
\eqn\Berryint
$$
is odd as shown in Appendix B.
The second term  in \Berryform\ is zero and the Berry phase is equal to
$$
\gamma = \pi. \eqn\eq
$$

  Recalling the calculation of ref [\KW], we find that the Berry phase
 is $\pi$ throughout the region $\Delta <0$.
  This confirms the continuity at $\Delta =-1$.
   It  will be different in the  region  $\Delta > 0$, which
we will find in the next subsections.

\section{ $0< \Delta < \cos \pi/3$ }

  The strategy for the computation is the same.
By using the permutation
$$
Q'_a=
\left\{
  \eqalign{Q_a+2,&\qquad ({\rm for}\ Q_a\neq M-1,M), \cr
           2,    &\qquad ({\rm for}\ Q_a=M-1), \cr
           1,    &\qquad ({\rm for}\ Q_a=M),}
\right.
\eqn\eq
$$
the relations \IIpa--\IIpd\ are expressed as
$$
p_{Q_a}^{(f)}=p_{Q'_a}^{(i)} + \pi(\delta_{Q'_a,1}+\delta_{Q'_a,2}),
\eqn\eq
$$
$$
\theta(p_{Q_b}^{(f)},p_{Q_a}^{(f)}) = \theta(p_{Q'_b}^{(i)},p_{Q'_a}^{(i)})
+2\pi\,\delta_{Q'_b,1}\delta_{Q'_a,2}
-2\pi\,\delta_{Q'_b,2}\delta_{Q'_a,1}.
\eqn\eq
$$
  Counting the change of the phase factors
$$
(-1)^{|Q|} = (-1)^{|Q'|}\times (-1),
\eqn\eq
$$
$$
{\rm exp} \left[ i\sum_{a=1}^Mx_ap_{Q_a}^{(f)} \right] =
{\rm exp} \left[ i\sum_{a=1}^Mx_ap_{Q'_a}^{(i)} \right],
\eqn\eq
$$
$$
  {\rm exp}\,
  \left[
    {i\over 2}\sum_{a<b}\theta(p_{Q_b}^{(f)},p_{Q_a}^{(f)})
    \epsilon(x_b-x_a)
  \right]
  = {\rm exp}\,
  \left[
    {i\over 2}\sum_{a<b}\theta(p_{Q'_b}^{(i)},p_{Q'_a}^{(i)})
    \epsilon(x_b-x_a)
  \right] \times (-1),
\eqn\eqeem
$$
we find, in the same way as we have derived Eqs. \esif\ ,
$$
|\chi(-2\pi)\bigr> = |\chi(2\pi)\bigr>.
$$
We conclude
$$
\gamma=0.
$$

\section{ $\cos \pi/3< \Delta < \cos \pi/4$ }

  The corresponding formulas in this case are
$$
Q'_a =
\left\{
\eqalign{ Q_3+3, \qquad &({\rm for}\ Q_a\neq M-2,M-1,M), \cr
          3,     \qquad &({\rm for}\ Q_a=M-2), \cr
          2,     \qquad &({\rm for}\ Q_a=M-1), \cr
          1,     \qquad &({\rm for}\ Q_a=M),}
\right. \eqn\eq
$$
$$
p_{Q_a}^{(f)}=p_{Q'_a}^{(i)}+2\pi\delta_{Q'_a,2},
\eqn\eq
$$
$$
\eqalign{
  \theta(p_{Q_b}^{(f)},p_{Q_a}^{(f)}) &= \theta(p_{Q'_b}^{(i)},p_{Q'_a}^{(i)})
  +2M\pi(\delta_{Q'_b,1}\delta_{Q'_a,2}-\delta_{Q'_b,2}\delta_{Q'_a,1})\cr
  &+ 2(M-2)\pi
  (\delta_{Q'_b,2}\delta_{Q'_a,3}-\delta_{Q'_b,3}\delta_{Q'_a,2})\cr
  &+2\pi
  (\delta_{Q'_b,1}\delta_{Q'_a,3}-\delta_{Q'_b,3}\delta_{Q'_a,1})\cr
  &+2\pi
  \left\{
    \delta_{Q'_b,2}\sum_{k=4}^M\delta_{Q'_a,k}-
    \sum_{k=4}^M\delta_{Q'_b,k}\delta_{Q'_a,2}
  \right\},}
\eqn\eq
$$
 and
$$
(-1)^{|Q|} = (-1)^{|Q'|}\times (-1)^{(M-1)+(M-2)+(M-3)},
\eqn\eq
$$
$$
{\rm exp}\,\left[ i\sum_{a=1}^Mx_ap_{Q_a}^{(f)}  \right] =
{\rm exp}\,\left[ i\sum_{a=1}^Mx_ap_{Q'_a}^{(i)} \right] \times (-1),
\eqn\eq
$$
$$
 {\rm e}^{{i\over 2}\sum_{a<b}\theta(p_{Q_b}^{(f)},p_{Q_a}^{(f)})
    \epsilon(x_b-x_a)}
= {\rm e}^{{i\over 2}\sum_{a<b}\theta(p_{Q'_b}^{(i)},p_{Q'_a}^{(i)})
    \epsilon(x_b-x_a)}\times (-1)^{M+(M-2)+1+(M-3)}.
\eqn\eq
$$
We conclude
$$
\gamma=\pi.
\eqn\eq
$$

\section{ $\cos (\pi / n)  \leq \Delta  \leq   \cos (\pi/ n+1)$ }

  By using the relations we  have found in  $\S 2.3.$,  we   compute
  the Bery phase in this regime  in the same way as we did
  in   $n=2, n=3$.  The final answer is
$$
  \gamma =0 ~~~~{\rm for }~ n~ {\rm even}
$$
$$
  \gamma = \pi ~~~~{\rm for}~ n~ {\rm odd}.
$$

\section{Discussion}

   Finally we would like to argue that the Berry phase we have computed
  can be used as a measure of  the statistics of the  quasiparticle
  ( or the bound state) involved in the process.  The above calculation
  shows
$$
  \gamma/ \pi = m ~~~~~ {\rm mod}~2.
\eqn\stat
$$
  for the case in which the $m$-string goes around the loop in the
 momentum/rapidity  plane.  This quantity  $\gamma/ \pi ~~{\rm mod}~2$
 measures  the monodromy property  in the $p$ space of the particle involved.
  We propose to use this as a definition of statistics.
   To make our discussion  a little more concrete,
 let us consider the case $\Delta =-1$.
 At $\Phi = 2\pi$, the largest rapidity ( or one-string) goes to the
  edge of the Brillouin zone ( or  infinity in the  $\lambda$ plane.)
   This root ( or the rapidity) decouples from the rest of the roots
  and  the system of $M$  equations eq. \trans reduces   that of
  $M-1$ equations.  According to the description of  [\FT], this is
  a two-spinon state over the vacuum.  Eq. \stat  provides
 $1$  for a creation   and a subsequent annihilation of
 two spinons, which implies  $1/2$ for a spinon.

\endpage

\chapter{Acknowledgements}
  We thank Vladimir Korepin for valuable discussion on this subject.
  We  are also grateful to  Yasuhiro Akutsu, Kiyoshi Higashijima,
 Keiji Kikkawa and V. P. Nair for insightful remarks.

\chapter{Note Added}
   After the completion of this paper, we became aware  that the formation
 and the collision of the two-string roots as a function of the coupling
 and the twist angle is  mentioned by Alcaraz, Barber and Batchelor
 in  [\ABBAN]. We thank Professor Batchelor for informing us of this reference.

\endpage

\Appendix{A}

In this appendix we examine  the case  $M=2,N=4$.
The boundary conditions \trans\ are
$$
4p_1 - \theta (p_2,p_1) = -\pi + \Phi ,
\eqn\atransa$$
$$
4p_2 - \theta (p_2,p_1) = +\pi + \Phi ,
\eqn\atransb$$
while the phase shift
$$
\theta(p_2,p_1) = 2 {\rm arctan}
\left[
  { {\Delta{\rm sin}[(p_2-p_1)/2]}\over
    {{\rm cos}[(p_2+p_1)/2] - \Delta{\rm cos}[(p_2-p_1)/2] } }
\right].
\eqn\Aps
$$
 In what follows, we solve explicitly
this system of equations.
The sum $\atransa+\atransb$ leads to the relation
$$
p_2+p_1 = {\Phi \over 2} .
\eqn\Asumm
$$
So the phase shift \Aps\ becomes
$$
\theta(p_2,p_1) = 2 {\rm arctan}
\left[
  { {\Delta{\rm sin}[(p_2-p_1)/2]}\over
    {{\rm cos}[\Phi / 4] - \Delta{\rm cos}[(p_2-p_1)/2] } }
\right].
\eqn\apsb
$$
{}From difference $\atransb-\atransa$ we obtain
 another formula for the phase-shift.
$$
\theta (p_2,p_1) = \pi -2 (p_2 - p_1) .
\eqn\apsc
$$
These formulas of the phase shift \apsb\ and \apsc\ lead to the equation on the
relative momentum
$$
2{\rm arctan} \left[ {\Delta {\rm sin}q \over {\rm cos}(\Phi / 4) - \Delta {\rm
cos} q} \right] = \pi - 4q
$$
where $q \equiv (p_2 - p_1)/2$.
This is expressed as  a relation
$$
2 \cos {\Phi \over 4} {\cos}^2 q - \Delta \cos q - \cos { \Phi \over 4} = 0 .
\eqn\Aseq
$$
When $ \cos (\Phi / 4) = 0 $, the equation becomes
$$
\cos q = 0.
$$
When $ \cos (\Phi / 4) \not= 0$, the roots of the equation \Aseq\ are
$$
\cos q = { \Delta \pm \sqrt{{\Delta}^2+8{\cos}^2 (\Phi / 4)} \over  4 \cos
(\Phi / 4)}.
\eqn\Asols
$$
In the case $|\cos q| \leq 1$, $p_1$ and $p_2$ are real number.
In the case of $|\cos q| >1$, $p_1$ and $p_2$ are complex number with
${p_1}^{\ast} = p_2$.
This set of the complex momenta is the two-string.
When $\cos (\Phi / 4) $ goes to zero, the relative momentum becomes
$$
\eqalign{
\lim_{\cos(\Phi /4) \rightarrow +0} \cos q^{(+)}
\equiv \lim_{\cos(\Phi /4) \rightarrow 0} {\Delta + \sqrt{ {\Delta}^2 + 8
{\cos}^2 (\Phi / 4) } \over  4 \cos ( \Phi / 4)}
\cr
= \Biggl{ \{ } \eqalign {   &+ \infty \ \ \ {\rm for} \ \Delta > 0, \cr  & 0 \
\  \  \  \  \ \ \ {\rm for} \ \  \Delta < 0,}
}
$$
$$
\eqalign{
\lim_{\cos(\Phi /4) \rightarrow +0} \cos q^{(-)}
\equiv \lim_{\cos(\Phi /4) \rightarrow 0} {\Delta - \sqrt{ {\Delta}^2 + 8
{\cos}^2 (\Phi / 4) } \over  4 \cos ( \Phi / 4)}
\cr
= \Biggl{ \{ } \eqalign {   &0 \ \ \ \ \ \ \ \ {\rm for} \ \Delta > 0, \cr  & +
\infty \  \  \  {\rm for} \ \  \Delta < 0.}
}
$$
 From \Asumm\ and \Asols, $p_1$ and $p_2$ are determined as
$$
\eqalign{
p_1 &= {\Phi \over 4} - {\rm arccos} \biggl{[}
{\Delta \pm \sqrt{ {\Delta}^2 + 8 {\cos}^2 (\Phi / 4)  } \over 4 \cos (\Phi /
4) }
\biggr{]} , \cr
p_2 &= {\Phi \over 4} + {\rm arccos} \biggl{[}
{\Delta \pm \sqrt{ {\Delta}^2 + 8 {\cos}^2 (\Phi / 4)  } \over 4 \cos (\Phi /
4) }
\biggr{]}.
}
$$

The wave function is determined from the momenta as follows.
In the present system the wave function \Bethest\ with \Bethewf\ is
$$
\eqalign{
| \chi \bigr> &= \sum_{x_1} \sum_{x_2} \epsilon(x_2-x_1) \sum_Q (-1)^{|Q|}
\exp (i x_1 p_{Q_1} + i x_2 p_{Q_2}) \cr
& \ \ \times \exp ({i \over 2} \theta (p_{Q_2},p_{Q_1}) \epsilon (x_2-x_1))
\sigma_{x_1}^- \sigma_{x_2}^- |\uparrow\bigr>
}
$$
The wave function is rewritten as
$$
\eqalign{| \chi \bigr> &= \sum_{ \{ x_1,x_2|x_2>x_1 \} }  \bigr{ \{ } \exp (i
x_1 p_1 + i x_2 p_2 + {i \over 2} \theta (p_1,p_2)) \cr & \ \ \ \ \  - \exp (i
x_1 p_2 + i x_2 p_1 - {i \over 2} \theta (p_1,p_2)) \bigl{ \} } \sigma_{x_1}^-
\sigma_{x_2}^- |\uparrow\bigr> \cr
&= 4i \sum_{\{ x_1,x_2|x_2>x_1 \} } \exp \bigr{(} i {p_1+p_2 \over 2} (x_2+x_1)
\bigl{)}
\cr
&\ \ \ \times \sin \bigr{(} i {p_2-p_1 \over 2}(x_2-x_1) + {1 \over 2} \theta
(p_1, p_2) \bigl{)} \sigma_{x_1}^- \sigma_{x_2}^- |\uparrow\bigr>
}
$$
By using \Asumm\ and \apsc, the wave function becomes
$$
| \chi \bigr> = 4i \sum_{ \{ x_1,x_2|x_2>x_1 \} } \exp \bigl{(} i {\Phi \over
4} (x_2+x_1) \bigr{)} \cos \bigl{(} { p_2-p_1 \over 2} (x_2-x_1-2) \bigr{)}
\sigma_{x_1}^- \sigma_{x_2}^- |\uparrow\bigr> .
\eqn\Awf
$$
We take the coordinate as
$$
\{ x \} = \{ -{3 \over 2}, {1 \over 2},{1 \over 2},{3 \over 2}. \}
$$
The wave function is described in terms of the relative momentum;
$$
\eqalign{
| \chi \bigr> &= 4i \bigl{]}
\exp \bigl{(} -i {\Phi \over 4} \bigr{)} |-{3 \over 2}, {1 \over 2} \bigr>
+\exp \bigl{(} +i {\Phi \over 4} \bigr{)} |-{1 \over 2}, {3 \over 2} \bigr> \cr
& \ \ \ + i\cos {p_2-p_1 \over 2} \bigl{ \{ }
\exp \bigl{(} -i {\Phi \over 4} \bigr{)} |-{3 \over 2}, -{1 \over 2} \bigr>
+|-{1 \over 2}, {1 \over 2} \bigr>
+\exp \bigl{(} +i {\Phi \over 4} \bigr{)} |{1 \over 2}, {3\over 2} \bigr>
+|{3\over 2}, -{3 \over 2} \bigr> \bigr{ \} }
\bigr{]} .}
$$
We get the explicit formula of the wave function by substituting \Asols\ into
this.

The eigenenergy is calculated from \Asumm\ and \Asols\
$$
E = \Delta \mp \sqrt{{\Delta}^2 + 8 {\cos}^2 (\Phi / 4)}.
$$
The energy, relative momentum and the momenta of the ground state are
$$
E^{(0)} = \Delta - \sqrt{{\Delta}^2 + 8 {\cos}^2 (\Phi / 4)}
$$
$$
\cos q^{(0)} =  {\Delta + \sqrt{{\Delta}^2 + 8 {\cos}^2 (\Phi / 4)} \over 4
\cos {\Phi \over 4}},
$$
$$
\eqalign{
p_1^{(0)} = {\Phi \over 2} - {\rm arccos}\left[ {\Delta + \sqrt{{\Delta}^2 + 8
{\cos}^2 (\Phi / 4)} \over 4 \cos {\Phi \over 4}} \right], \cr
p_2^{(0)} = {\Phi \over 2} + {\rm arccos}\left[ {\Delta + \sqrt{{\Delta}^2 + 8
{\cos}^2 (\Phi / 4)} \over 4 \cos {\Phi \over 4}} \right].
}
$$

Let us consider the case $ \Delta <0 $. In this case the momentum
 is real number because of
$$
| \cos q^{(0)} | = \left|{\Delta + \sqrt{{\Delta}^2 + 8 {\cos}^2 (\Phi / 4)}
\over 4 \cos {\Phi \over 4}}\right| \leq 1
$$
The motion of the momentum in the  process is as follows.
When $\Phi =0$, the momenta are
$$
\eqalign{
p_1^{(i)} = - {\rm arccos} \left[ { \Delta + \sqrt{ {\Delta}^2 +8} \over 4}
\right], \cr
p_2^{(i)} = + {\rm arccos} \left[ { \Delta + \sqrt{ {\Delta}^2 +8} \over 4}
\right] ,
}
\eqn\Api
$$
thus the momentum is arranged symmetric $p_1^{(i)} = -p_2^{(i)}$.
When $\Phi = 2 \pi$, $\cos q^{(0)}= 0$, so the momenta are
$$
p_1=0, \ \ \ p_2 = \pi.
$$
When $\Phi = 4 \pi$, the state comes back to the ground state. The set of
momenta $\{ p_1^f, p_2^f \}$ at $\Phi = 4 \pi$ is equal to the set of momenta
$\{ p_1^{(i)}, p_2^{(i)} \}$ at $\Phi =0$ up to $2\pi$;
$$
\eqalign{
p_1^f &= {4 \pi \over 4} - \arccos \left[ - {\Delta + \sqrt{{\Delta}^2 + 8}
\over 4} \right] \cr
&= \pi - \left( \pi - {\rm arccos} \left[ - {\Delta + \sqrt{{\Delta}^2 +8}
\over 4} \right] \right) \cr
&= -{\rm arccos} \left[ - {\Delta + \sqrt{{\Delta}^2 + 8} \over 4} \right] \cr
&= p_2^{(i)} ,
}
$$
$$
\eqalign{
p_2^f &= {4 \pi \over 4} + \arccos \left[ - {\Delta + \sqrt{{\Delta}^2 +8}
\over 4} \right] \cr
&= \pi + \left( \pi - {\rm arccos} \left[ - {\Delta + \sqrt{{\Delta}^2 + 8}
\over 4} \right] \right) \cr
&= 2\pi+ p_1^{(i)} ,
}
\eqn\Apf
$$
Our assertion on the motion of the rapidities in Sec.4-3 is checked by
\Api-\Apf.

Next we consider the case $\Delta >0$.
We take the region $-2\pi < \Phi < 2\pi$ for convenience.
Note the relation
$$
\eqalign{
|\cos q^{(0)}| &= \left|{\Delta + \sqrt{{\Delta}^2 + 8 {\cos}^2 (\Phi / 4)}
\over 4 \cos {\Phi \over 4}}\right| \cr
& \biggl{ \{ }\eqalign{
\leq 1 & \ \ \ {\rm for} \ \ -2(\pi - \mu) \leq \Phi \leq 2(\pi-\mu), \cr
  >1  &   \ \ \ {\rm for} \ \ -2\pi < \Phi < -2(\pi - \mu), \  2(\pi -
\mu)<\Phi
 <2 \pi,
}}
$$
we check that the two-string is formed at $\Phi = 2(\pi-\mu)$.
In the two-string the momenta is complex number with $p_1^{\ast}=p_2$ which is
equivalent to the center of rapidities $(\lambda_1, \lambda_2)$ is on the
$i\pi$ line (see \moment). When $\Phi$ goes to 2$\pi$, the relative momentum
become $q^{(0)}\to i\infty$ because of $\cos q^{(0)}\to\infty$, so the momentum
becomes
$$
\eqalign{
p^f_1 &= {\pi\over 2}-i\infty,\cr
p^f_2 &= {\pi\over 2}+i\infty .
}
$$
{}From this and  Eq. \moment, we find the rapidities arrived at the singular
point
$$
\eqalign{
\lambda^f_1&= i(2\pi-\mu),\cr
\lambda^f_2&= i\mu .
}
$$
In the same way, we obtain the momenta and rapidities at $\Phi=-2\pi$ as
$$
\eqalign{
p^i_1 &=-{\pi\over 2}+i\infty,\cr
p^i_2 &=-{\pi\over 2}-i\infty ,
}
$$
$$
\eqalign{
\lambda^i_1&= i\mu,\cr
\lambda^i_2&= i(2\pi-\mu).
}
$$
So, we get the relations
$$
\lambda^f_2=\lambda^i_1,\qquad \lambda^f_1=\lambda^i_2,
$$
$$
p^f_2=p^i_1+\pi, \qquad p^f_1=p^i_2+\pi.
$$
They agree with \IIpe-\IIpa.

The set $\{p^f_j\}$ is not equivalent to the set $\{p^i_j\}$ up to $2\pi$.
However the initial state and final state are the same state.
The relative momentum $\cos q^{(0)}=\infty$ at $\Phi=\pm 2\pi$,
thus the wave function \Awf\ becomes
$$
\eqalign{
|\chi(\pm2\pi)\bigr> &\propto \Bigl{\{}
\exp ({-i{\pm 2\pi\over 4}})
|-{3\over 2},{1\over 2}\big>+
\exp ({i{\pm 2\pi\over 4}})
|-{1\over 2},{3\over 2}\big>\Bigr{\}
}\cr
&+i^2\infty\times\left(
\exp ( {-i{\pm 2\pi\over 2}})
|-{3\over 2},-{1\over 2}\big>+
|-{1\over 2},{1\over 2}\big>+
\exp({i{\pm 2\pi\over 2}})
|{1\over 2},{3\over 2}\big>+
|{3\over 2},-{3\over 2}\big>
\right),
}
$$
the normalized wave function is
$$
\eqalign{
|\chi(\pm2\pi)\bigr>&=-{1\over 2}\Big[
\exp({-i{\pm 2\pi\over 2}})
|-{3\over 2},-{1\over 2}\big>+
|-{1\over 2},{1\over 2}\big>+
\exp({i{\pm 2\pi\over 2}})
|{1\over 2},{3\over 2}\big>+
|{3\over 2},-{3\over 2}\big>
\Big]\cr
&=-{1\over 2}\Big[-|-{3\over 2},-{1\over 2}\big>+
|-{1\over 2},{1\over 2}\big>-|{1\over 2},{3\over 2}\big>+
|{3\over 2},-{3\over 2}\big>
\Big].
}
$$
Therefore we obtain $|\chi(2\pi)\bigr>=|\chi(-2\pi)\bigr>$.

Finally we examine the energy spectrum by the explicit diagonalization
of the Hamiltonian \Hferm\ with the twisted boundary condition
which is equivalent to the Hamiltonian
$$
\eqalign{
H&=\sum_j\big(
e^{i{\Phi \over N}}\tilde C^{\dag}_j\tilde C_{j+1}+
e^{-i{\Phi \over N}}\tilde C^{\dag}_{j+1}\tilde C_j\big)
-2\Delta\sum_j\big(\tilde C^{\dag}_j\tilde C_j-{1\over 2}\big)
\big(\tilde C^{\dag}_{j+1}\tilde C_{j+1}-{1\over 2}\big),
}
\eqn\AHf$$
with boundary condition
$$
\tilde C_{j+N}=\tilde C_j, \qquad \tilde C^{\dag}_{j+N}=\tilde C^{\dag}_j,
$$
where
$$
\tilde C_j=e^{i{j \over N}\Phi}C_j,\qquad
\tilde C^{\dag}_j=e^{-i{j \over N}\Phi}C^{\dag}_j.
$$
The diagonalization of the Hamiltonian \AHf\ for $N=4$,
$M=2$ system leads to the energy spectrum
$$
E=\Delta\pm\sqrt{\Delta^2+8\cos^2({\Phi\over 4})},\
\Delta\pm\sqrt{\Delta^2+8\sin^2({\Phi\over 4})},\  0,\  0.
$$
We plot the spectrum as a function of $\Phi$ in the case $\Delta < 0$,
$\Delta = 0$ and $\Delta > 0$ in Sec. 4-4.

\endpage


\Appendix{B}

The set $\{p_j^{(i)}\}$ and the set $\{p_j^{(f)}\}$ are not equivalent
even up to $2\pi$, as seen in \IIpa\ and \IIpb.
In spite of this, the initial state $(\Phi=-2\pi)$ and the final state
$(\Phi=2\pi)$ are the same.
To show this, we see that the wave function $|\chi (-2 \pi)>$ of the final
state (at $\Phi = 2\pi$) is invariant under the set of the transformations
$p_M^{(f)} \rightarrow p_M^{(f)} + \pi$ and $p_{M-1}^{(f)} \rightarrow
p_{M-1}^{(f)} + \pi$.
When $\Phi=2\pi-\epsilon\ (0<\epsilon\ll 1)$, the rapidities of two
strings are $\lambda_M^{(f)} \simeq i(2\pi-\mu)+\delta$ and
$\lambda_{M-1}^{(f)}\simeq i\mu+\delta$, where $0<\delta\ll 1$.
Substituting this into \moment, we obtain
$$
p_M^{(f)}     \simeq -i\Lambda + {\pi\over 2}, \ \ \
p_{M-1}^{(f)} \simeq  i\Lambda + {\pi\over 2},
\eqn\Bpf
$$
where $\Lambda \equiv {\rm ln} (2 \sin \mu / \delta) \gg 1$; $\Lambda$ goes to
infinity for
$\epsilon\rightarrow 0$.
The phase shift $\theta(p_M^{(f)},p_{M-1}^{(f)})$ is determined from
the twisted boundary conditions (see \trans),
$$
2Mp_M^{(f)}
+\theta(p_M^{(f)},p_{M-1}^{(f)})+\sum_{k=1}^{M-2}\theta(p_M^{(f)},p_k^{(f)})
= 2\pi\left( {M-1\over 2} \right) + 2\pi,
\eqn\eq
$$
$$
2Mp_{M-1}^{(f)}+\theta(p_{M-1}^{(f)},p_M^{(f)})+\sum_{k=1}^{M-2}\theta(p_{M-1}^{(f)},p_k^{(f)})
= 2\pi\left( {M-3\over 2} \right) + 2\pi.
\eqn\eq
$$
These lead to the relation
$$
\eqalign{&2M(p_M^{(f)}-p_{M-1}^{(f)})+2\theta(p_M^{(f)},p_{M-1}^{(f)})
          + \sum_{k=1}^{M-2}
          \left(
            \theta(p_M^{(f)},p_k^{(f)})-\theta(p_{M-1}^{(f)},p_k^{(f)})
         \right) \cr
         =& 2\pi.}
\eqn\Btrans
$$
{}From \Bpf\ and \Btrans\ the phase shift becomes
$$
\theta(p_M^{(f)}, p_{M-1}^{(f)}) \simeq 2iM\Lambda + \pi - {1\over 2}
\sum_{k=1}^{M-2}(\theta(p_M^{(f)},p_k^{(f)})-\theta(p_{M-1}^{(f)},p_k^{(f)})).
\eqn\eq
$$
If we substitute this into the Bethe's wave function \Bethest\ and \Bethewf,
the $\Lambda$ dependence of the wave function of the final state (at $\Phi = 2
\pi $) becomes
$$
\eqalign{
& |\chi(2\pi)\bigr> \cr
= & \sum_{x_1,\cdots,x_M}
  \prod_{a<b}\epsilon(x_b-x_a)
\sum_Q(-1)^{|Q|}{\rm e}^{i\sum_a x_{Q_a} p_a +
  {i\over 2}\sum_{a<b}\theta(p_b,p_a)\epsilon(x_{Q_b}-x_{Q_a})}
\prod_j\sigma_{x_j}^{-}|\uparrow\bigr>  \cr
= & \sum_{x_1,\cdots,x_M}
\prod_{a<b}\epsilon(x_b-x_a)\sum_Q(-1)^{|Q|}
{\rm exp}
\left[
\Lambda \big{(} x_{Q_M}-x_{Q_{M-1}}- M \epsilon(x_{Q_M}-x_{Q_{M-1}}) \big{)}
\right] \cr
& \ \ \ \ \times ({\rm independent\ part\ of}\ \Lambda)
\prod_j\sigma_{x_j}^{-}|\uparrow\bigr>. } \eqn\wf
$$
Considering the range of the coordinate, it is shown that
the leading terms ($ \simeq e^{\Lambda (M-1)}$) of the wave function \wf\
satisfy the condition $x_{Q_M}-x_{Q_{M-1}} = -1$ or $2M-1$:
$$
\eqalign{
(x_{Q_M}, x_{Q_{M-1}}) &=
(-{N-1\over 2},-{N-3\over 2}),
(-{N-3\over 2},-{N-5\over 2}),
\ldots \cr
& \quad\ldots,
({N-5\over 2},{N-3\over 2}),
({N-3\over 2},{N-1\over 2}),
({N-1\over 2},-{N-1\over 2}).}
\eqn\xQi
$$
These $x_{Q_M}$ and $x_{Q_{M-1}}$ are adjacent to each other on the ring.
When $\epsilon \rightarrow 0$ ($\Lambda \rightarrow \infty$), the leading terms
go to infinity faster than the other terms. Therefore, after the normalization
of the wave function, the leading terms remain and the other terms vanish.
The wave function at $\Phi = 2 \pi$ ($\epsilon \rightarrow 0$) is the sum of
the terms which satisfy the condition \xQi.
Under the set of the transformations $p_M^{(f)}\rightarrow p_M^{(f)}+\pi$ and
$p_{M-1}^{(f)}\rightarrow p_{M-1}^{(f)}+\pi$,
the phase factor ${\rm exp}i\sum x_{Q_a}p_a$
in the wave function is transformed as
$$
{\rm e}^{i\sum x_{Q_a}p_a}
\rightarrow
{\rm e}^{i\sum x_{Q_a}p_a}\times
{\rm e}^{i\pi(x_{Q_M}+x_{Q_{M-1}})}.
\eqn\pfac
$$
As $x_{Q_M}+x_{Q_{M-1}}$ is even as seen in \xQi, the phase factor in \pfac\
becomes
unity:
$$
{\rm e}^{i\pi(x_{Q_M}+x_{Q_{M-1}})}=1.
\eqn\eq
$$
Thus the wave function at $\Phi=2\pi$ does not change under the set of
transformations $p_M^{(f)}\rightarrow p_M^{(f)}+\pi$ and
$p_{M-1}^{(f)}\rightarrow p_{M-1}^{(f)}+\pi$.
Therefore the initial state and the final state are equivalent through
the above set of the transformations.

\endpage

\Appendix{C}

In this appendix, we show that the function \Berryint\ is odd function of
$\Phi$.
We start from the twisted boundary condition
$$
Np_j(\Phi) + \sum_{k=1(\neq j)}^M\theta(p_j(\Phi),p_k(\Phi)) = 2\pi I_j + \Phi
,
\eqn\Ctbc
$$
where $N=2M$, $I_j={2j-M-1  \over 2}$ ($j=1,2, \cdots , M$) and
$$
\theta(p_j,p_k) = 2 {\rm arctan}
\left[
  { {\Delta{\rm sin}[(p_j-p_k)/2]}\over
    {{\rm cos}[(p_j+p_k)/2] - \Delta{\rm cos}[(p_j-p_k)/2] } }
\right].
$$
By noting $\theta(p_j(\Phi)$, $p_k(\Phi))=-\theta(p_k(\Phi),p_j(\Phi))$ and
$I_j=-I_{M-j+1}$, \Ctbc$\times (-1)$ is written as
$$
N(-p_j(\Phi)) + \sum_{j=1,k\neq j}^M\theta(-p_j(\Phi),-p_k(\Phi)) = 2\pi
I_{M+1-j} - \Phi .
$$
Here we rewrite $n+1-j$ to $j$;
$$
N(-p_{M+1-j}(\Phi)) + \sum_{k=1(\neq
M+1-j)}^M\theta(-p_{M+1-j}(\Phi),-p_k(\Phi)) = 2\pi I_j - \Phi .
$$
Compare this and boundary condition at $(-\Phi)$
$$
Np_j(-\Phi) + \sum_{k=1(\neq j)}^M\theta(p_j(-\Phi),p_k(-\Phi)) = 2\pi I_j -
\Phi ,
$$
we obtain
$$
p_j(-\Phi) = -p_{M+1-j}(\Phi) .
$$
Using this relation and $\epsilon(x_b-x_a) = - \epsilon ((-x_a)-(-x_b)) $ the
wave function \Bethewf\ becomes
$$
\eqalign{
& \chi(x_1, \cdots,x_M|\Phi) \cr
= & \prod_{a<b}\epsilon(x_b-x_a)
\sum_Q(-1)^{|Q|}{\rm e}^{i\sum x_ap_{Q_a}(\Phi)+
{i\over 2}\sum_{a<b}\theta(p_{Q_b}(\Phi),p_{Q_a}(\Phi))\epsilon(x_b-x_a)
} \cr
=& (-1)^{M(M-1)/2} \prod_{a<b}\epsilon((-x_b)-(-x_a))
\sum_Q(-1)^{|Q|}{\rm e}^{i\sum (-x_a)p_{M+1-Q_a}(-\Phi)} \cr
 & \ \ \ \times {\rm exp}
\left[
{i\over
2}\sum_{a<b}\theta(p_{M+1-Q_b}(-\Phi),p_{M+1-Q_a}(-\Phi))\epsilon((-x_b)-(-x_a))
\right].}
$$
Here we put $Q'_a \equiv M+1-Q_a$, then this wave function is expressed as
$$
\eqalign{
& \chi(x_1,\cdots,x_M|\Phi) \cr
=& \prod_{a<b}\epsilon((-x_b)-(-x_a))
\sum_{Q'}(-1)^{|Q'|}{\rm e}^{i\sum (-x_a)p_{Q'_a}(-\Phi)}
{\rm e}^{{i\over
2}\sum_{a<b}\theta(p_{Q'_b}(-\Phi),p_{Q'_a}(-\Phi))\epsilon((-x_b)-(-x_a))}\cr
=& \chi(-x_1, \cdots,-x_M|-\Phi),
}
$$
where we used $(-1)^{|Q|}  = (-1)^{|Q'|} \times (-1)^{M(M-1)/2}$.
Therefore we get the relation
$$
\chi(x_1,  \ldots,x_M|\Phi) = \chi(-x_1, \ldots,-x_M|-\Phi).
$$
{}From this relation  and $ \{ x_j \}  =  \{- x_j \} $  it is shown that the
function \Berryint\ is odd function of $\Phi$;
$$
\eqalign{
& \sum_{\{x_i\}}
{\chi(x_1,\ldots,x_M|\Phi)^*
{{\partial \over \partial\Phi}}
\chi(x_1,\ldots,x_M|\Phi) \over
\|\chi(x_1 \ldots , x_M|\Phi) \|^2} \cr
=& \sum_{\{x_i\}}
{\chi(-x_1,\ldots,-x_M|-\Phi)^*
(-{{\partial \over \partial(-\Phi)}})
\chi(-x_1,\ldots,-x_M|-\Phi) \over
\|\chi(-x_1, \ldots ,-x_M|- \Phi) \|^2} \cr
=& -\sum_{\{x_i\}}
{\chi(x_1,\ldots,x_M|-\Phi)^*
({{\partial \over \partial(-\Phi)}})
\chi(x_1,\ldots,x_M|-\Phi) \over
\|\chi(x_1, \ldots , x_M|-\Phi) \|^2}.
}
$$

\endpage

\refout

\end